\documentclass[3p,times,review]{elsarticle}

\usepackage[figuresright]{rotating}

\usepackage{lipsum}
\makeatletter
\def\ps@pprintTitle{%
 \let\@oddhead\@empty
 \let\@evenhead\@empty
 \def\@oddfoot{\centerline{\thepage}}%
 \let\@evenfoot\@oddfoot}
\makeatother

\usepackage{stmaryrd,bm,amsmath}
\usepackage{dsfont}
\usepackage{amsbsy,amsmath,amsthm,latexsym,amssymb,mathtools}
\usepackage{graphicx,stmaryrd,sectsty}
\usepackage{setspace}
\usepackage[font=small,labelfont=bf]{caption}
\usepackage{epstopdf}
\usepackage{verbatim}
\usepackage{graphicx}
\usepackage{caption}
\usepackage{subcaption}
\usepackage{float}
\usepackage[toc,page]{appendix}
\usepackage{tikz}
\usepackage{enumitem}
\usepackage[utf8]{inputenc}

\newcommand{\B}{{Ba\v zant}}
\newcommand{\bc}{\begin{center}}
\newcommand{\ec}{\end{center}}
\newcommand{\bfr}{\begin{flushright}}
\newcommand{\efr}{\end{flushright}}

\newcommand{\be}{\begin{enumerate}}
\newcommand{\ee}{\end{enumerate}}
\newcommand{\bi}{\begin{itemize}}
\newcommand{\ei}{\end{itemize}}
\newcommand{\bd}{\begin{description}}
\newcommand{\ed}{\end{description}}
\newcommand{\beq}{\begin{equation}}
\newcommand{\eeq}{\end{equation}}
\newcommand{\bea}{\begin{eqnarray}}
\newcommand{\eea}{\end{eqnarray}}

\newcommand{\bfi}{\begin{figure}}
\newcommand{\efi}{\end{figure}}
\newcommand{\bay}{\begin{array}{l}}
\newcommand{\eay}{\end{array}}

\newcommand{\cref}[1]{(\ref{#1})}   

\sectionfont{\normalfont\fontsize{14}{19}\bfseries}
\subsectionfont{\normalfont\fontsize{12}{17}\bfseries}
\subsubsectionfont{\normalfont\itshape\fontsize{12}{17}\selectfont}

\begin{document}

\begin{frontmatter}

\title{Meso-scale Finite Element Modeling of Alkali-Silica-Reaction}

\author[l1,l2]{Roozbeh Rezakhani  \footnote{Corresponding author. \\ E-mail address: rrezakhani@u.northwestern.edu, jean-francois.molinari@epfl.ch}}
\author[l2]{Emil Gallyamov}
\author[l2]{Jean-Fran\c cois Molinari}

\address[l1]{Department of Mechanical Engineering and Material Science, Duke University, Durham, NC, USA.}
\address[l2]{Civil Engineering Institute, Materials Science and Engineering Institute, Ecole Polytechnique Fédérale de Lausanne (EPFL), Lausanne,
Switzerland.}

\begin{abstract}
The alkali-silica reaction (ASR) in concrete is a chemical reaction, which produces an expansive product, generally called “ASR gel”, and causes cracking and damage in concrete over time. Affecting numerous infrastructures all around the world, ASR has been the topic of much research over the past decades. In spite of that, many aspects of this reaction are still unknown. In this numerical-investigation paper, a three-dimensional concrete meso-structure model is simulated using the finite-element method. Coarse aggregates, cement paste, and ASR gel are explicitly represented. A temperature dependent eigen-strain is applied on the simulated ASR gel pockets to capture their expansive behavior. This applies pressure on the surrounding aggregates and the cement paste, leading to cracks initiation and propagation. Free expansion of concrete specimens due to ASR is modeled and validated using experimental data. Influence of different key factors on damage generation in aggregates and paste and macroscopic expansion are discussed.
\end{abstract}


\end{frontmatter}
\section{Introduction}	
Structural effects due to the Alkali-Silica Reaction (ASR) were first noticed in the State of California, and Thomas Stanton from the California State Division of Highways was the one who reported ASR effect in 1940 \cite{stanton1940}. Stanton studied the expansion of mortar bars and showed that the type and the amount of the reactive silica existing in aggregates, the amount of alkali content in cement, and the level of moisture and temperature are among influential factors. Furthermore, he showed that the ASR induced expansion is negligible when the amount of alkali content in the cement paste is below a certain limit. Following Stanton’s work, abnormal cracking caused by ASR was diagnosed in several dams operated by the US Bureau of Reclamation, including the Parker Dam in Arizona (Meissner \cite{Meissner1941}). ASR is now recognized worldwide as a major cause of infrastructure deterioration and has been the topic of intense research investigation. 

ASR being a concern for large infrastructures, it is no surprise that the majority of theoretical and numerical models are macroscopic and aim to capture the consequences of ASR directly at the structural scale of concrete. In macroscopic models, concrete is simulated as a homogeneous isotropic material, which implies that the underlying meso-scale constituents, such as aggregates, cement paste, and ASR gel, are not taken into account explicitely. In these models, the macroscopic (continuum) constitutive laws are formulated either by fitting the available experimental data \cite{capra1998modeling, saouma2006constitutive} or through homogenization principles \cite{capra2003orthotropic, bangert2004chemo}, which account for the underlying concrete meso-structure only implicitly. Among the continuum based models available in the literature, we highlight Comi et al. \cite{comi2009chemo, comi2011anisotropic} who developed an orthotropic damage constitutive law accounting for the chemical and mechanical aspects of ASR in a consistent thermodynamic fashion. Poyet et al. \cite{poyet2007chemical} integrated the effect of temperature and humidity factors into the ASR process by developing a theory based on the transport of alkali and calcium ions in concrete. \B~et al. \cite{rahimi-1,rahimi-2} incorporated ASR effect into a microplane model by applying expansive pressure at random integration points and investigated the effect of ASR gel flow on the macroscopic expansion. Numerical modeling of ASR structural effects on concrete dams was performed by Léger et al. \cite{leger}, who proposed to relate different portions of the observed concrete expansion to various dominant factors. Regarding theoretical formulation for ASR kinetics, Larive \cite{Larive1997} proposed a sigmoidal function to describe ASR gel expansion in time by considering two characteristic time parameters and a final gel strain expansion. Temperature dependence of the gel expansion was formulated by Ulm et al. \cite{larive}, and Capra and Bournazel \cite{capra1998modeling} have included the effect of humidity into the reaction kinetics. Saouma and Perotti \cite{saouma2006constitutive} incorporated the effect of compressive stresses arguing that high value of hydrostatic stresses slow down the reaction process. However, this assumption contradicts the experimental evidences \cite{Larive1997,dunant2012effects} stating that ASR reaction kinetics do not depend on the applied stresses, which in fact affects ASR by influencing the orientation of cracks in concrete. The developed ASR expansion models have been implemented into various finite element software for the analysis of large structural systems, as in for instance the work of Fairbairn et al. \cite{fairb}.

Since macro-scale models do not take into account the root of the macroscopic deterioration of concrete at a lower scale, they must rely to some degree on phenomenological relationships between macroscopic expansion, damage, and advancement of the chemical reaction. Such models overlook important fine-scale details of ASR processes and, in turn, provide only limited insights about its origins and driving factors. This has triggered the development of micro- and meso-scale mechanical models. In most of the fine-scale ASR models, it is assumed that the ASR product forms and grows at the surface of the aggregates resulting in cracks in the interfacial zone between the aggregates and cement paste \cite{bazant2000mathematical, bazant2000fracture, suwito2002mathematical, multon2009chemo, charpin2012computational, charpin2014microporomechanics,rezakhani2019multiscale}. Comby-Peyrot et al. \cite{comby2009development} simulated ASR effect in a meso-scale concrete model through expanding all the aggregates and generating damage in paste. Alnaggar et al. \cite{alnaggar2017modeling, alnaggar2013lattice} proposed a discrete model simulating concrete through an ensemble of rigid particles, each representing aggregates and the surrounding cement paste. ASR was then simulated by imposing a normal opening on the interfaces between the particles leading to macroscopic expansion. However, it has been observed in laboratory experiments that ASR gel can form inside the aggregates \cite{leemann2013modulus}. Subsequently, upon gel expansion, cracks initiate and primarily develop inside the aggregates, which is followed by further propagation into the cement phase. In this regard, Dunant et al. \cite{dunant2009experimental, dunant2010micro} developed a two-dimensional finite element model simulating the concrete internal structure and ASR product, which was extended by Giorla et al. \cite{giorla2015influence} to investigate the effect of paste viscoelasticity. Cuba Ramos et al. \cite{ramos2018hpc} recently incorporated the main features of the aforementioned 2D model in a parallel computational framework.

In this paper, a viscoelastic three-dimensional meso-scale finite element framework is developed. A new formulation for ASR gel expansion is proposed, which takes into account the effect of ambient temperature and its variation on the rate of ASR. ASR free expansion of concrete is validated with laboratory experiments. It is shown that the main part of concrete macroscopic ASR expansion is due to the development of a crack network, while the expansion associated to elastic deformation is negligible. In addition, we discuss how the slow initial ASR expansion rate, which is observed in most laboratory measurements, is related to the elastic deformation and crack initiation phase. This is naturally captured by a model that assumes a linear ASR gel expansion in the initial growth phase, with no need to assume an ``S" type expansion curve, which considers a slow expansion kinetic in the beginning. It is shown that the macroscopic ASR expansion accelerates when the damage network further develops in aggregates and paste. Furthermore, the effect of different influential factors on ASR including ASR gel mechanical properties and growth rate, viscoelasticity, and temperature are investigated using the proposed model.

\begin{figure}[t!]
  \centering 
  \includegraphics[width=\textwidth]{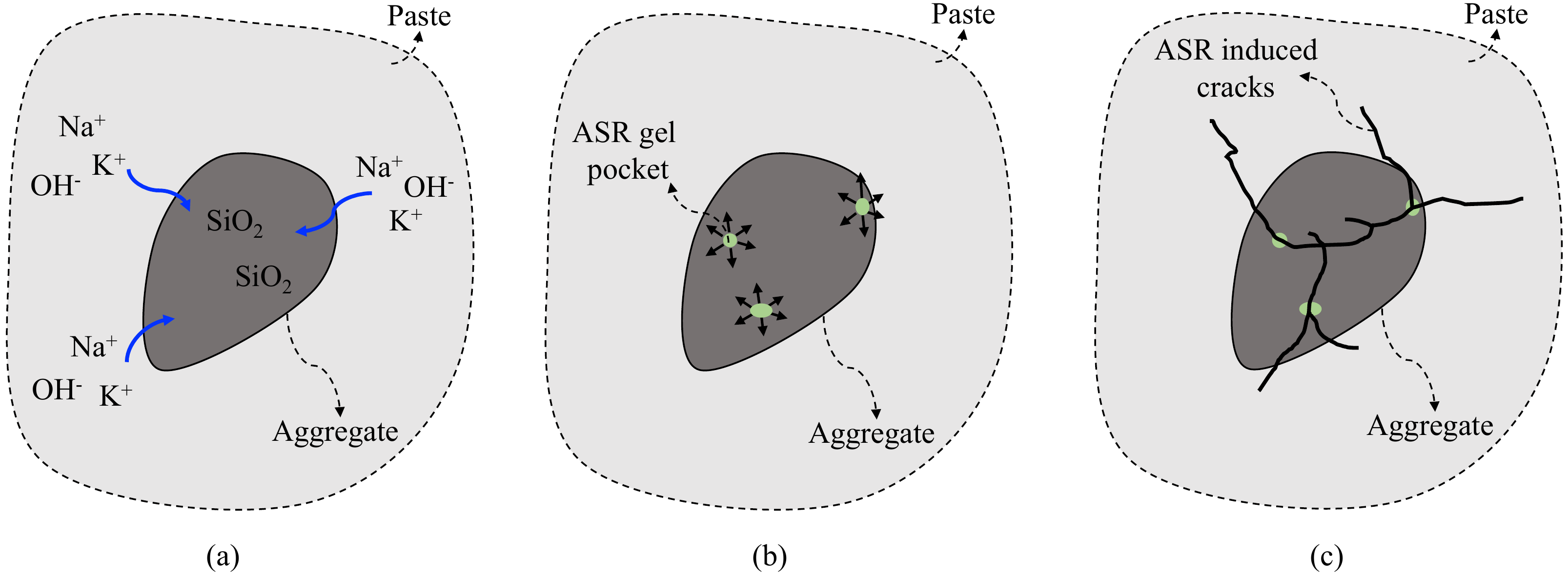}
  \caption{(a) A generic aggregate including silica placed inside the surrounding paste containing alkali ions and humidity. (b) ASR gel pockets generated inside the aggregate. (c) ASR induced cracks due to the expansion of ASR product.}
  \label{ASR-process}
\end{figure}

\section{ASR and meso-structural finite-element modeling}\label{FE-gen}
\begin{figure}[t!]
  \centering 
  \includegraphics[width=\textwidth]{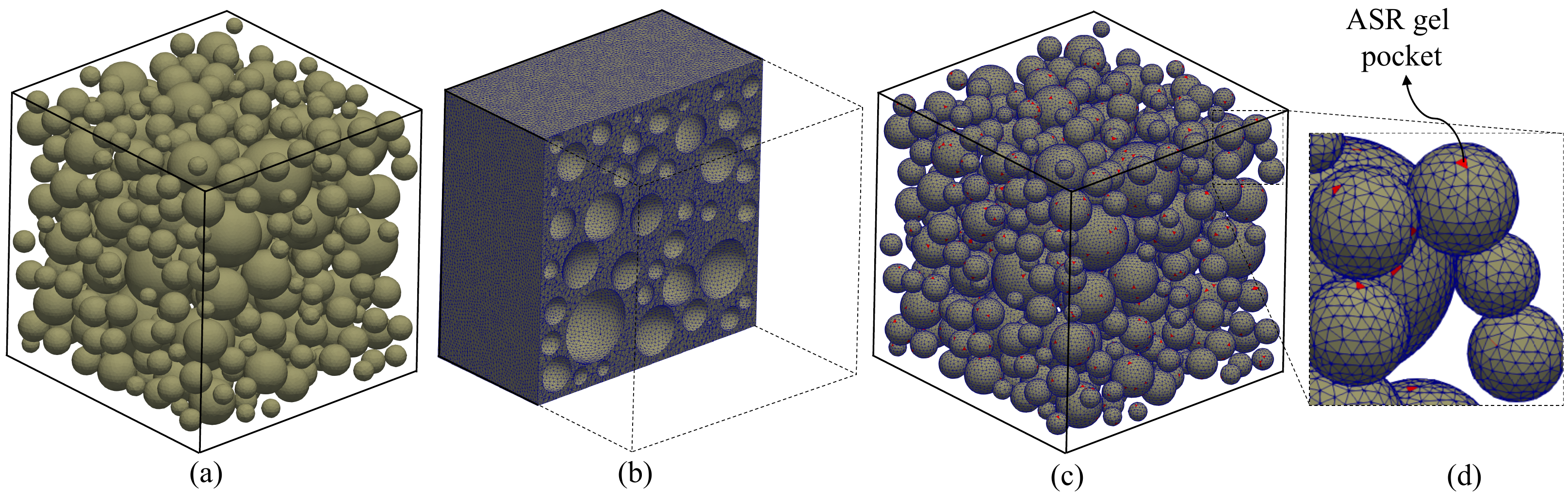}
  \caption{(a) A generic spherical aggregates distribution in a cubic sample. (b) Finite-element mesh discretization of the paste. (c) Finite element discretization of spherical aggregates and the ASR gels (gel elements are shown in red). (d) An enlarged view of discretized aggregate pieces and the ASR gels.}
  \label{2D-geometry}
\end{figure}

During ASR, different types of alkali ions, including K$^{+}$ and Na$^{+}$, existing in the cement paste permeate into concrete aggregates, and in the presence of water (H$_{2}$O) and hydroxide ions (OH$^{-}$), react with the silica located inside the aggregate pieces \cite{dron1992thermodynamic,chatterji1987studies}, see Figure \ref{ASR-process}a.  As the product of this reaction, ASR gel is produced, whose chemical composition varies because of different available alkali ions as well as aggregates mineralogies \cite{GLASSER1979515}. In addition, it has been confirmed in laboratory experiments that silica is randomly distributed in aggregate pieces in the form of discrete pockets and is not uniformly dispersed over the aggregates volume \cite{chatterji1986studies}. This indicates that the ASR gels can appear in an arbitrary pattern inside and on the surface of aggregates, see Figure \ref{ASR-process}b, which is an important aspect that needs to be addressed in the numerical simulations. Due to the expansive nature of the produced gel, cracks are initiated in aggregates and propagate into the cement phase, as shown in Figure \ref{ASR-process}c, which gives rise to a meso-scale crack network in concrete.
To simulate ASR, a meso-structural model is presented in this paper, which is based on a three-dimensional finite-element modeling of concrete internal structure. In this model, concrete aggregates, paste, and ASR gel pockets are explicitly simulated. It should be noted that this model analyzes the ASR problem after the ASR gels generation in aggregates and does not consider the alkali diffusion. The first step is to construct a concrete meso-structure by distributing aggregate pieces within a representative volume \cite{wriggers2006mesoscale}. In this method \cite{wriggers2006mesoscale}, spherical aggregate particles are first randomly distributed in the specimen, and the mortar paste is then considered to fill the space between the aggregates in the sample. To perform the random aggregates placing process, spherical particles are chosen from a specific size distribution function corresponding to a given sieve curve. In practice, a Fuller curve is often used to represent the grading of aggregate particles sizes, which is described by $P(d) = (d/d_{max})^{n_f}$, and is employed in the current algorithm. $P(d)$ is a cumulative distribution function defining relative aggregates weight passing through a sieve with the characteristic size of $d$. $d_{max}$ is the maximum aggregate size, and $n_f$ is the Fuller exponent. A minimum aggregate size $d_{min}$ is chosen to obtain the simulated aggregates volume fraction $v_{agg}=[1-P(d_{min})]v^0_{agg} = [1-(d_{min}/d_{max})^{n_f}] v^0_{agg}$, in which $v^0_{agg}$ is the aggregates volume fraction in the experimental concrete batch \cite{cusatis2011lattice}. The total volume of the simulated aggregates is then computed as $V_{agg} = v_{agg}V_0$, where $V_0$ is the concrete specimen volume. Until the intended aggregates volume fraction is met, aggregate particles are placed one by one into the volume box in a way that no overlapping occurs with the already placed particles. Figure \ref{2D-geometry}a presents a generic aggregates distribution generated by this algorithm in a cubic sample of size $70$ mm using $n_f = 0.5$; $d_{min}=4$ mm; $d_{max}=20$ mm; $v_{agg}=0.4$. The concrete meso-structure is then completed by considering cement paste in the spaces between the aggregates. Next, paste and aggregates are discretized by finite elements as depicted in Figure \ref{2D-geometry}b and c, respectively. To include ASR gel in the model, finite elements inside the aggregates are randomly converted to ASR gel until a certain percentage of reactive materials denoted by $\overline{V}_{ga} = V_{gel}/V_{agg}$ is met. $V_{gel}$ is the total volume of ASR product. The randomly distributed ASR gels are shown in Figure \ref{2D-geometry}c, and in a zoomed-view in Figure \ref{2D-geometry}d, by red finite elements, for which $\overline{V}_{ga} = 1.5\%$ is used. Isotropic volumetric eigen-strain is then applied on the generated gel elements to imitate their expansive behavior during alkali-silica chemical reaction and subsequent water absorption, which exerts pressure on the surrounding materials. This procedure is explained in Section \ref{ASR-gel-const} in more details. Note that these geometrical parameters will be varied in the following parametric studies. One should note that in the current model, the effect of the Interface Transition Zone (ITZ), the weak interfaces between each aggregate and surrounding paste, is not investigated. The best approach to study such effect is utilizing a cohesive element framework, in which cracking takes place on the interfaces between elastic finite elements. This aspect is out of the scope of this paper and will be investigated in future work.

\section{Material constitutive models} \label{Constitutive-models}
\subsection{Viscoelastic behavior of paste}
A solidifying Kelvin chain model \cite{bavzant2018creep} is adopted to simulate the viscoelastic response of paste in concrete, which is the root of different time-dependent characteristic responses of concrete, such as creep and shrinkage. A Kelvin chain consists of a single elastic spring with stiffness $E_0$ coupled in series with $N$ viscous units each consisting of an elastic spring with stiffness $E_\mu$ in parallel to a dashpot with viscosity $\eta_\mu$ ($\mu=1,2,..,N$). The compliance function of such model is given by
\begin{equation}\label{J-kelvin}
J(t,t') = \large \left[ \frac{1}{E_0} + \sum_{\mu=1}^N  \frac{1}{E_\mu}\bigg(1-e^{-(t-t')/\tau_\mu}\bigg) \large \right]
\end{equation}
in which $\tau_\mu = \eta_{\mu}/E_{\mu}$ is the retardation time of each viscous unit. $t'$ is the creep test start time and $t-t'$ is the load duration at any given time $t$. Having the compliance function of the kelvin model, the strain tensor at time $t$ generated by a general stress history can be written as
\begin{equation}\label{epsilon-kelvin}
\boldsymbol{\epsilon}(t) = \int_0^t J(t,t')~\bold{C}_\nu~\text{d}\boldsymbol{\sigma}(t')
\end{equation}
with $\bold{C}_\nu$ the elastic compliance matrix considering unit Young's modulus. In solidification theory, it is assumed that the stiffness and viscosity of the viscous units are time dependent, which is related to the deposition of a non-aging constituent. In this case, these parameters are formulated as $E_\mu(t) = E_\mu^{\infty}v(t)$ and $\eta_\mu(t) = \eta_\mu^{\infty}v(t)$, in which $v(t)$ is the function that characterizes the volume of the solidified material. $E_\mu^{\infty}$ and $\eta_\mu^{\infty}$ are constant material parameters. $v(t)$ is defined as $v(t) = 1/(\alpha+\sqrt{\lambda_0/t})$ by Ba{\v{z}}ant et al. \cite{bavzant1997microprestress}, in which $\alpha$ is a constant parameter and $\lambda_0=1$ day. One should consider that $t$ is the real physical time, which should be calculated based on the simulation final time and numerical time step through appropriate time mapping as explained in Appendix \ref{visc-explicit}. Numerical integration of Equation \ref{epsilon-kelvin} and the calculation of creep strain increment at each time step are presented in Appendix \ref{visc-explicit}, and more details can be found in \cite{bavzant2018creep}. 

\subsection{Damage constitutive model for aggregates and paste} 
Nonlinear behavior of both aggregates and paste are simulated using the Mazars' damage model \cite{mazars1986description, mazars1989continuum}. This damage model aims at simulating the reduction of the material stiffness by considering the evolution of a scalar damage variable $D$. The damage variable $D$ varies from $0$ corresponding to a no-damage state, or intact material, to $1$ representing the completely damaged material state with zero stiffness. To formulate the variation of D, an effective strain is defined as follows
\begin{equation}\label{effective-strain}
\tilde{\epsilon} = \sqrt{\sum_{I=1}^3 \left\langle \epsilon_I \right\rangle _{+}^2}
\end{equation}
where $\left\langle  \right\rangle_{+}$ operator denotes the Macaulay brackets for which $\left\langle x \right\rangle_{+} = \text{max} \left\lbrace 0,x \right\rbrace$. $\epsilon_I$ are the principal strain components, in which $I$=1 to 3 in a three-dimensional setting. Equation \ref{effective-strain} implies that only positive principal strains are considered in effective strain calculation, and the effect of compressive strains is ignored. The general damage criteria is defined as
\begin{equation}\label{damage-initiation}
f(\epsilon, D) = \tilde{\epsilon} - K(D) = 0
\end{equation}
in which $K(D)$, at any point in the material domain, is equal to the largest value of the effective strain $ \tilde{\epsilon}$ experienced by the material at that point. Initially, $K(D) = k_0$, and $k_0$ is a material parameter representing the initial damage threshold. In Mazars' damage model, damage scalar $D$ is decomposed into two damage variables $D_t$ and $D_c$ describing damage generated in tension and  compression, respectively. The total damage is formulated as the weighted sum of $D_t$ and $D_c$ as follows
\begin{equation}\label{damage-decompos}
D = \alpha_t D_t + \alpha_c D_c
\end{equation}
in which 
\begin{equation}\label{Dt-Dc}
D_{t,c} = 1 - \frac{k_0(1-A_{t,c})}{\tilde{\epsilon}} - A_{t,c} \exp[-B_{t,c}(\tilde{\epsilon}-k_0)]
\end{equation}
where ($A_t$, $B_t$) and ($A_c$, $B_c$) are constant parameters governing damage evolution in tension and compression, respectively. The weight coefficients $\alpha_t$ and $\alpha_c$ are formulated as follows
\begin{equation}\label{alpha_tc}
\alpha_t = \sum_{I=1}^3 \frac{\epsilon_{I}^t \left\langle \epsilon_{I} \right\rangle_{+}}{\tilde{\epsilon}^2}; \hspace{3 mm} \alpha_c = \sum_{I=1}^3 \frac{\epsilon_{I}^c \left\langle \epsilon_{I} \right\rangle_{+}}{\tilde{\epsilon}^2}
\end{equation}
where $\epsilon_{I}^t$ and $\epsilon_{I}^c$ are the $I$th positive and negative principal strains associated to the corresponding parts of the stress tensor in principal coordinates. Therefore, $\alpha_t$ and $\alpha_c$ are related through $\alpha_t + \alpha_c = 1$. Calculation details of the Mazars damage law during each time step are presented in Appendix \ref{mazars-explicit}. One of the main drawbacks of the classical damage models is the dependence of the dissipated energy on the finite element size. Several approaches exist in literature to overcome this issue, among which the crack band model proposed by Ba\v zant \cite{bavzant1983crack} is selected in this research to resolve the mesh dependency problem associated with strain-softening behavior. This issue is addressed in more details in Section \ref{Numerical results}. Finally, having the damage variable $D$ calculated for the current strain state of the material $\boldsymbol{\epsilon}$, the stress tensor $\boldsymbol{\sigma}$ can be calculated through
\begin{equation}\label{strs-strn}
\boldsymbol{\sigma} = (1-D)E\bold{D_\nu}\boldsymbol{\epsilon}
\end{equation}
in which $\bold{D_\nu}$ is the dimensionless elastic stiffness matrix corresponding to a unit value of Young’s modulus. One should consider that in the calculations of finite elements in the paste phase, $\boldsymbol{\epsilon}$ is the elastic strain tensor, which is calculated by subtracting the creep strain tensor from the strain tensor at the current time step. This corresponds to the assumption of series coupling between damage and viscoelasticity, in which creep behavior takes place in the solid between the microcracks \cite{bavzant2018creep}. Several other strategies are also available in the literature, among which the coupling of viscoelasticity and damage through a constant parameter as in Bottoni et al. \cite{Bottoni-creepdamage}.

\subsection{Gel mechanical and expansive behavior}\label{ASR-gel-const}
In this study, ASR gel is considered to behave elastically, characterized by the elastic modulus $E_{gel}$ and the Poisson's ratio $\nu_{gel}$. Recently, Gholizadeh Vayghan et al. \cite{gholizadeh2019characterization} studied viscoelastic properties of ASR gel with different chemical compositions, which will be included in our model in future work. ASR gel is assumed to expand in isotropic fashion, with volumetric expansion imposed as an eigen-strain tensor $\boldsymbol{\epsilon_{gel}}$ on the gel finite element as
\begin{equation}\label{eig-str}
\boldsymbol{\epsilon_{gel}} = \epsilon_{gel}(t)~\textbf{I}
\end{equation}
in which $\textbf{I}$ is the identity tensor, and $\epsilon_{gel}(t)$ is the amount of ASR gel expansion at any given time during the analysis. The rate of the chemical reaction, which defines the rate of ASR gel formation and subsequent expansion, is temperature dependent, and this dependency is described by the Arrhenius law. It describes the reaction rate based on the surrounding temperature through
\begin{equation}\label{eq:arr}
k=Ke^{-E_a/RT}
\end{equation}
where $k$ is the reaction rate, $K$ is a constant, $E_a$ is the activation energy reported in literature \citep{BenHaHa2006a}, $R$ is the gas constant, and $T$ is the temperature. Applied to the ASR case, the increase in the ambient temperature accelerates the reaction and leads to production of larger amount of gel, which in turn leads to higher local strain level.
\begin{figure}[t!]
  \centering 
  \includegraphics[width=\textwidth]{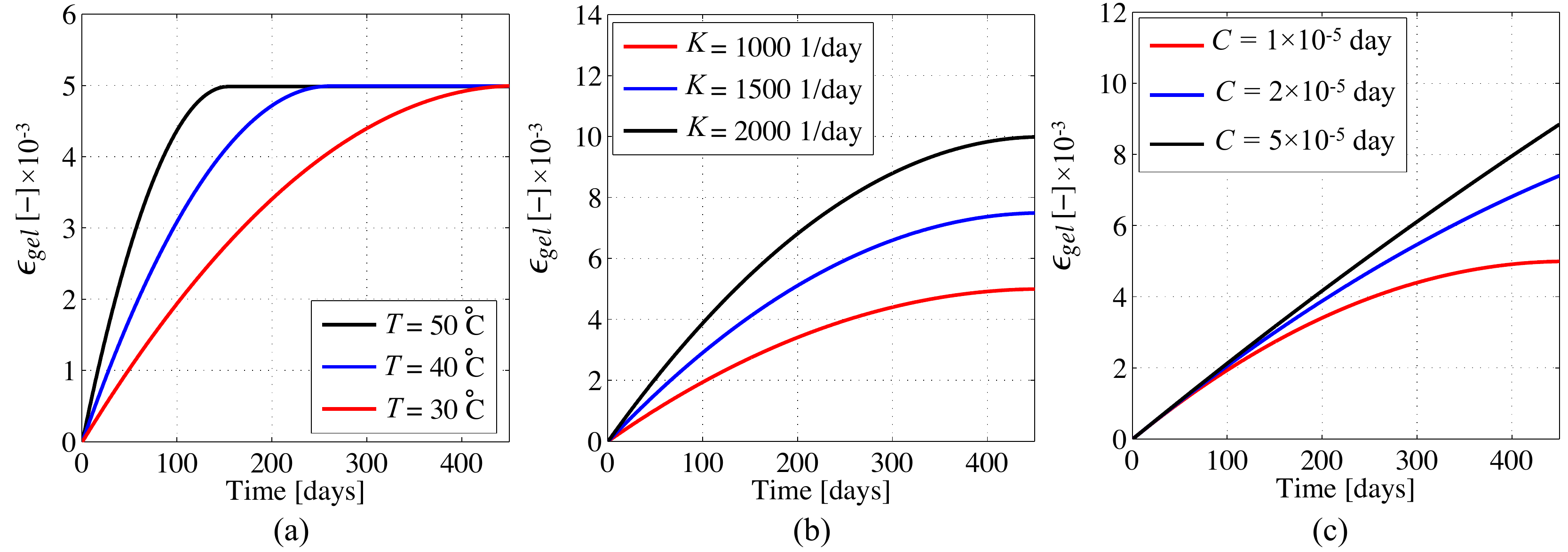}
  \caption{Evolution of gel strain for different (a) temperature $T$ values with $C=1\times10^{-5}$ day and $K=1000$ 1/day, (b) $K$ values with $C=1\times10^{-5}$ day and $T = 30^\circ$C, and (c) $C$ values with $K=1000$ 1/day and $T = 30^\circ$C. }
  \label{gel-exp}
\end{figure}

In order to quantitatively link the ambient temperature with the loading produced by the ASR gel, several phenomena have to be taken into consideration. Among them are the ASR gel formation rate, redistribution of the gel within neighbouring pores and micro-cracks, change of its properties with time and the surrounding environment. All these phenomena take place at the nanometer scale, and there is still limited understanding of them. Leemann et al. \cite{Leemann2019,Leemann2016} recently showed that the primary ASR product starts to accumulate between the mineral grains within reactive concrete aggregates. The scale of such reaction is finer than the meso-scale, which is the smallest scale that our model can take into account. Therefore, the expansion that is applied in our model at a single finite element represents the homogenized expansion of a single gel pocket and a certain size of the encompassing aggregate domain, rather then expansion of the gel itself. However in this paper, it is referred to as the ``ASR gel expansion" for the sake of brevity. A direct link between the ambient temperature and the local expansion rate is proposed, which is based on the assumption that the local expansive strain is proportional to the total amount of produced gel. This should also hold for the time derivative of the above mentioned phenomena, which means $k\propto\dot{\epsilon}_{gel}$, where $k$ is defined in Equation \ref{eq:arr}. Under this assumption, Arrhenius law is applicable for computing the local expansion rate of an ASR gel and the surrounding aggregate domain.
 
In most of the laboratory experiments, the measured macroscopic expansion of specimens plateaus after a certain time \cite{Larive1997,Multon2006a,Gautam2016a}. This can be attributed to different reasons such as a finite source of reactants or the formation of large interconnected cracks network, in which the produced gel starts to flow, reducing the induced pressure on the surrounding material. In addition, some researchers associate the plateau of expansion curve to alkali leaching, which is not fully prevented in the experiments \cite{multon2016multi}. In order to account for the decay in gel expansion rate, the following modification to the Arrhenius law is proposed
\begin{equation} \label{strain_rate}
\dot{\epsilon}_{gel} = K A(t, T) e^{-E_{a}/RT} 
\end{equation}
\begin{equation} 
A(t, T) = \max{\left[\frac{1}{C} \left(C - \int_0^t e^{-E_{a}/RT} dt \right), 0\right]}
\label{strain_rate-A}
\end{equation}

In the above equations, $\dot{\epsilon}_{gel}$ is the expansion rate of the ASR gel; $A(t, T)$ is the dimensionless parameter representing remaining amount of reactants linearly decreasing from $1$ to $0$; $T$ is the temperature at a specific time $t$; $C$ is the saturation constant, which is of time unit and represents the reaction duration of a single gel pocket. The effect of temperature $T$, and parameters $K$ and $C$ on the cumulative local expansion $\epsilon_{gel}$ are shown in Figure \ref{gel-exp}. It can be seen that the parameter $C$ affects the plateau value and can convert an exponential gel expansion to a linear one over a certain period of time. While $C$ does not affect the slope of the initial linear part, increasing $K$ directly increases the initial reaction rate without shifting the saturation time. It should be noted that the effect of relative humidity on gel expansion is not included in this model, and the laboratory experiments selected for simulation were performed under constant relative humidity close to 100\% or saturation condition.

\begin{figure}[t!]
  \centering 
  \includegraphics[width=0.9\textwidth]{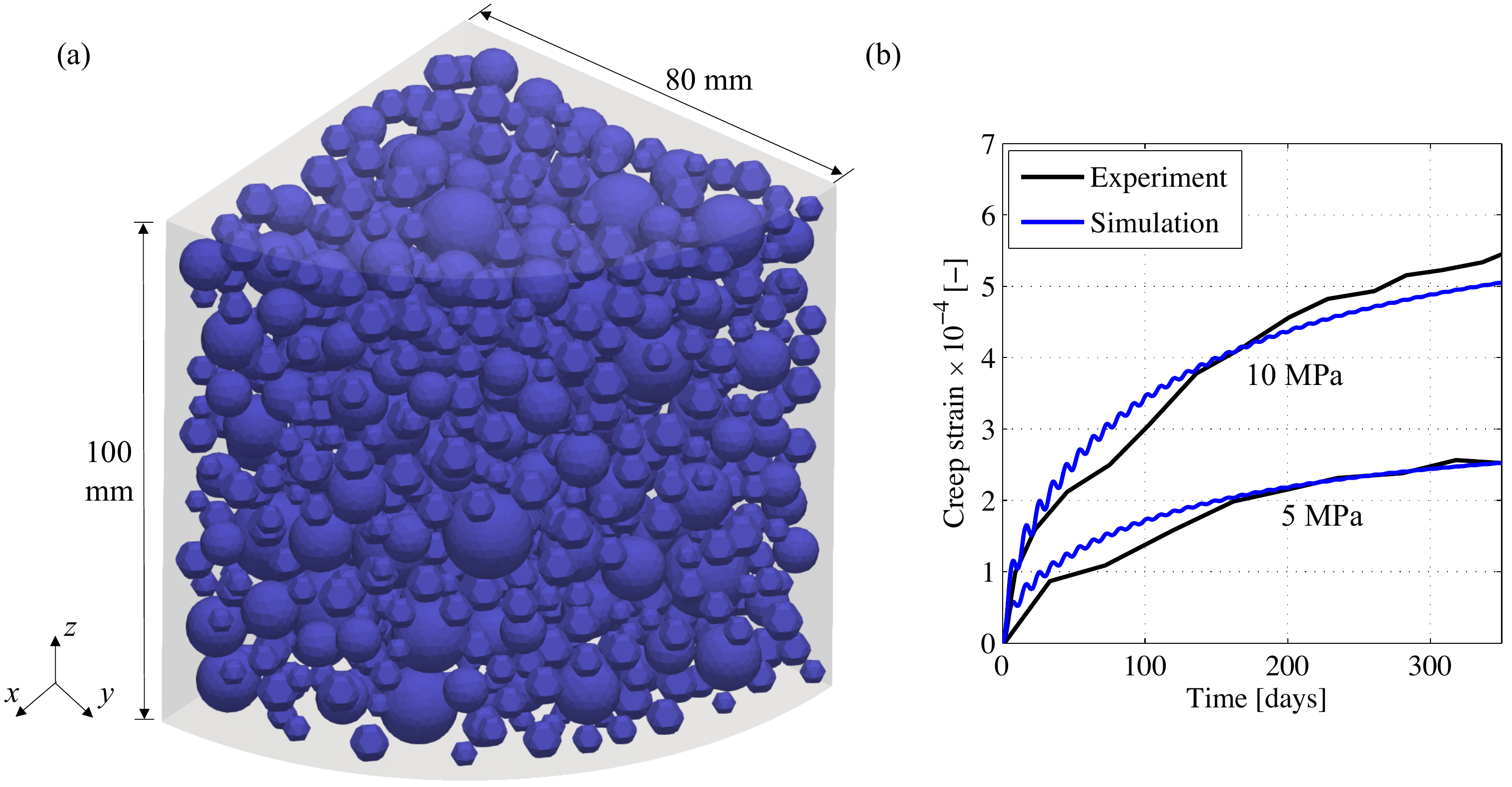}
  \caption{(a) Quarter cylinder simulated to perform the creep test. (b) Creep strain obtained from the numerical simulation in comparison with experimental data reported by Dunant et al. \cite{dunant2012effects} on cylinders under 5 and 10 MPa compressive uniaxial stresses.}
  \label{creep-test}
\end{figure}

\begin{table}[t!]
\centering
\caption{Paste viscoelastic parameters with three viscous units.}
\label{visc-paramters}
\begin{tabular}{c||c||c|c||c|c||c|c||c}
\hline
Material & $E_0$ [GPa] & $E_1$ [GPa] &  $\tau_1$ [Days] & $E_2$ [GPa] &  $\tau_2$ [Days] & $E_3$ [GPa] &  $\tau_3$  [Days] & $\alpha$ [-] \\ \hline
Paste & 20  & 12 & 5 & 8 & 50 & 0.7 & 300 & 0.001  \\ 
\hline
\end{tabular}
\end{table}

\section{Numerical results} \label{Numerical results}
\subsection{Creep experiment simulation}\label{creep-sim}
Paste viscoelastic parameters are calibrated using the creep data presented by Dunant et al. \cite{dunant2012effects}. In the experiments, cylindrical specimens with radius $R=80$ mm and height $H=335$ mm are tested under uniaxial compressive stresses of 5 and 10 MPa for 350 days. In the finite-element simulation, a quarter of the cylinder along with 100 mm height of the sample is modeled to reduce the computational cost as shown in Figure \ref{creep-test}a. The following boundary conditions are applied to reproduce the pure uniaxial deformation: $u_x = 0$ on $yz({x=0})$ plane, $u_y = 0$ on $xz({y=0})$ plane, $u_z = 0$ on $xy({z=0})$ plane, $\sigma_z = -5$ MPa on $xy({z=H})$ plane. One should consider that since concrete is a heterogeneous material with large inclusion sizes, the assumption on the symmetry of the solution and the applied displacement boundary conditions on the symmetry planes are not locally satisfied in a fully simulated sample. However, given that the characteristic size of the sample is approximately five times the maximum aggregate size, this simplifying assumption of symmetry should not have considerable effect on the global creep behavior of the sample. Aggregates size distribution of the concrete samples in experiment are reported in Reference \cite{dunant2012effects}, which shows presence of fine aggregate pieces with diameter in the range of 0-4 mm. However, since each aggregate is discretized with finite elements in the numerical simulation, the radius of the smallest aggregate size taken into account defines an approximate higher-bound limit on the finite element size. Therefore, to reduce the computational cost, $d_{min}=4$ mm is chosen as the minimum aggregate size in the simulations. The maximum aggregate size in the simulations is $d_{max}=16$ mm, which corresponds to the reported experimental value. The whole sample is discretized with uniform 2 mm size linear tetrahedral finite elements. The following elastic properties are employed in the simulation: aggregates Young's modulus $E_{agg} = 60$ GPa, taken from Ben Haha \cite{BenHaHa2006a}; its Poisson's ratio $\nu_{agg} = 0.2$; and paste Poisson's ratio $\nu_{paste} = 0.2$. The numerical simulations start by an initial static step, through which the axial stress is applied on the sample. Next, an explicit dynamic analysis is carried out to compute the evolution of strain in the specimen during time. The viscoelastic properties of paste are calibrated with respect to the experimental data for the case of 5 MPa axial stress and are reported in Table \ref{visc-paramters}. Since in this numerical simulation, aggregates smaller than 4 mm are not modeled and are homogenized into the paste phase between simulated aggregates, elastic stiffness of the single spring in paste material model is set to $E_0=20$ GPa, which corresponds to mortar Young's modulus adapted from Boumiz et al. \cite{boumiz1996mechanical}. Three viscous units with characteristic times of $\tau_{\mu} = 5, 50, 300$ days are considered. The calibrated parameters are then used for the simulation with 10 MPa applied axial stress to predict the model response. The creep strain is measured in the numerical simulation and is plotted in Figure \ref{creep-test}b, which shows a good agreement with respect to the experimental results for both calibrated and predicted cases. The minor oscillations in the numerical results are due to the explicit dynamic algorithm employed to solve the time dependent creep problem. It should be noted that since both creep tests are performed on mature concrete at approximately the same age, solidifying aspect of the Kelvin chain model is not crucial here.

\begin{table}[h!]
\centering
\caption{Mazars damage model parameters for aggregates and paste.}
\label{mazars-paramters}
\begin{tabular}{c||c|c|c|c|c}
\hline
Material & $k_0$ & $A_t$ &  $B_t$ & $A_c$ &  $B_c$ \\ \hline
Paste & 2$\times 10^{-4}$  & 0.65 & 3100 & 1.00 & 2300 \\ \hline
Aggregate  & 1.67$\times 10^{-4}$  & 0.65 & 3550 & 1.20 & 1800 \\
\hline
\end{tabular}
\end{table}

\begin{figure}[h!]
  \centering 
  \includegraphics[width=\textwidth]{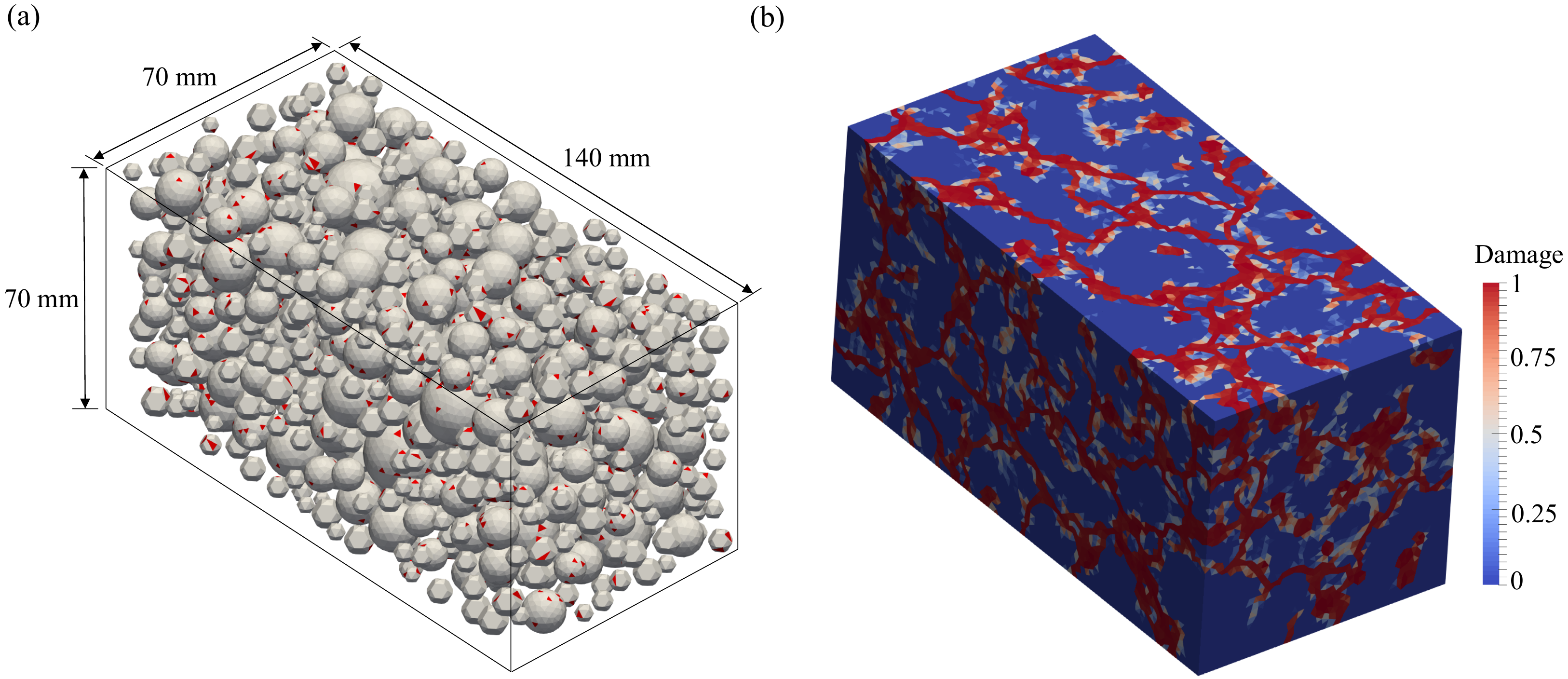}
  \caption{(a) Aggregate particles in the prismatic sample with red elements representing ASR gels (the ones that are placed on the surface of the aggregates are only visible in this figure). (b) Damage contour of the sample after ASR free expansion simulation.}
  \label{ASRfree-geom-contour}
\end{figure}

\begin{figure}[h!]
  \centering 
  \includegraphics[width=0.8\textwidth]{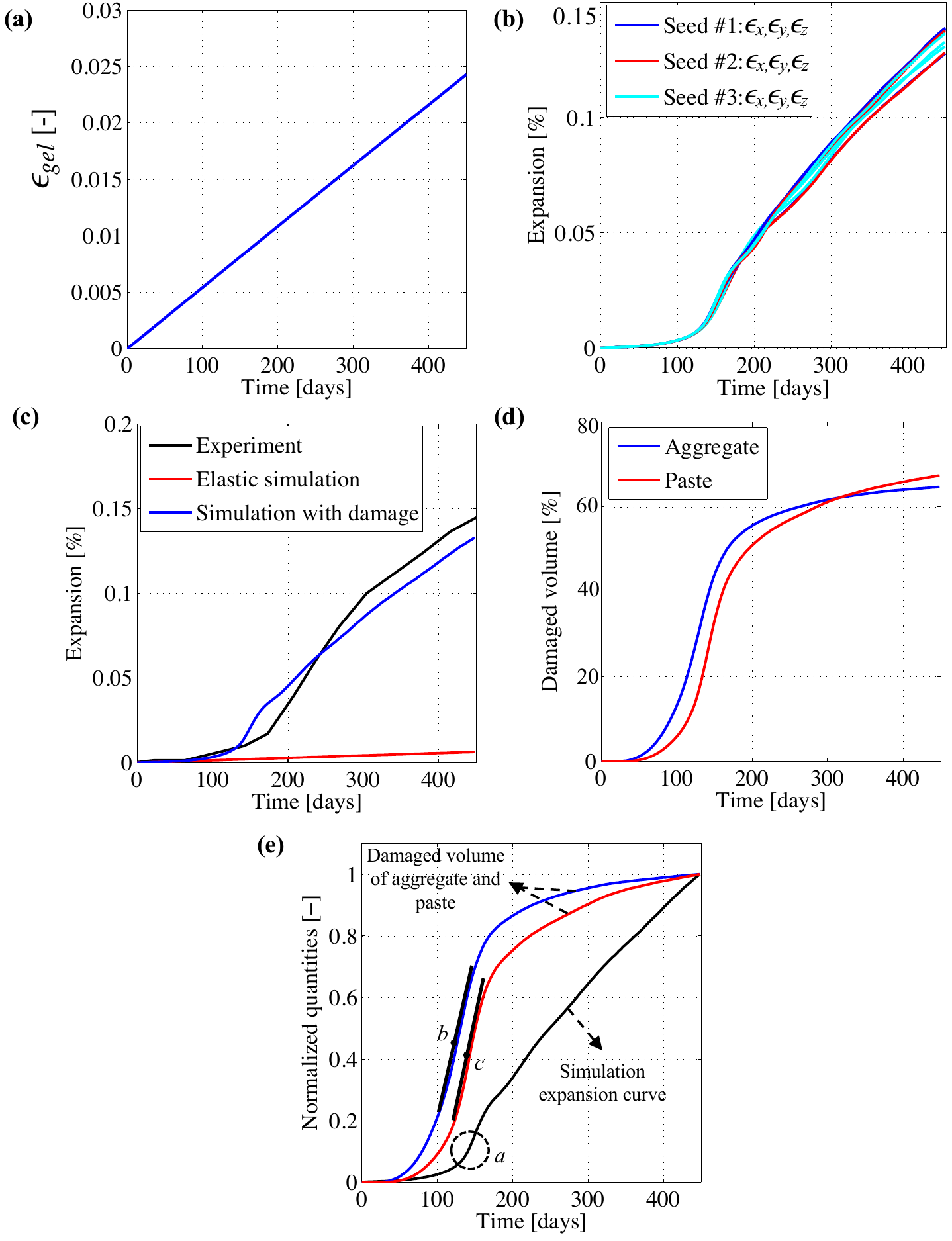}
  \caption{(a) ASR gel expansion evolution in time. (b) ASR expansion of the prisms simulated with three different inernal aggregate realizitations measured in three directions $x$, $y$, and $z$. (c) Axial expansion of the prism due to ASR obtained from numerical simulations and the experimental data by Dunant et al. \cite{dunant2009experimental}. (d) Evolution of damaged volume percentage in aggregates and paste. (e) Normalized representation of the simulation axial expansion curve presented in plot (c) and the damaged volume curves of aggregates and paste presented in plot (d).}
  \label{ASRfree-results}
\end{figure}

\subsection{ASR free expansion simulation}\label{ASR-free}
Dunant et al. \cite{dunant2009experimental} reported ASR free expansion experiments conducted on prismatic concrete samples with the same concrete mix design used for the creep test presented in Section \ref{creep-sim}, except for the maximum aggregate size that is increased to $d_{max}=20$ mm. Dimensions of prismatic samples are $70\times70\times280$ mm, which is reduced to $70\times70\times140$ mm in the numerical simulations to reduce computational cost. The whole sample is uniformly discretized by $2$ mm size linear tetrahedral finite elements. It should be considered that the smallest simulated aggregates with $d_{min}=4$ mm are coarsely discretized, which affects the intended volume fraction of aggregates in the simulation, parameter $v_{agg}$ in Section \ref{FE-gen}. In the current example, using the equation presented in Section \ref{FE-gen} with $n_f=0.5$, the target value for $v_{agg}$ is equal to 40\%, using which the aggregates are placed inside the sample. The volume fraction of the simulated aggregates are then computed using the volume of the finite elements placed in aggregates phase of the discretized sample, which is equal to 37.8\% for the generated prism in this example. This difference is considerably reduced in the simulation of the same sample discretized with 1 mm finite element size presented later in this section, for which the volume fraction of aggregates computed using the volume of finite elements is equal to 39.4\%.

Mazars damage model constitutive parameters are calibrated for paste and aggregate to match the tensile strength of $f^{\prime}_{t,paste}=4$ MPa for paste and $f^{\prime}_{t,agg}=10$ MPa for aggregates, based on the experimetal data reported by Chen et al. \cite{chen2013influence} and Ben Haha \cite{BenHaHa2006a}, respectively. Calibrated parameters of the Mazars damage model are presented in Table \ref{mazars-paramters}. \B's~crack band model \cite{bavzant1983crack} states that the area under the stress-strain curve multiplied by the corresponding finite element size is equal to the dissipated energy in that element. Therefore, to dissipate the desired fracture energy of $G_{agg} = 150$ N/m, based on Ben Haha \cite{BenHaHa2006a} for aggregates, and $G_{paste} = 60$ N/m, taken from Xu and Zhu \cite{xu2009experimental} for paste, in a $2$ mm finite element, the ultimate fracturing strain of 2\%, after which damage variable $D=1$, is used for both material phases. It is worth repeating here that the adopted values for mechanical properties of paste correspond to the reported experimental data for mortar since aggregates of diameter less that 4 mm are homogenized into the paste phase in the finite element model. In the experiments, free expansion of the concrete samples are measured for $450$ days, while alkali leaching is partially prevented to exclude its effect on the strain measurements. Generated aggregate pieces in the prismatic sample are shown in Figure \ref{ASRfree-geom-contour}a, in which the red elements represent the ASR gels randomly distributed in aggregates volume. The ratio of the ASR gel volume to the aggregates volume, denoted by $\overline{V}_{ga} = V_{gel}/V_{agg}$, is set to $2.5$\% in the current example. The effect of this parameter will be investigated in Section \ref{ASRgel-perc}. The gel expansion related parameters are: $C = 50\times 10^{-5}$ day, $E_a = 43500$ J/mol, $K = 2500$ 1/day, $R = 8.1344$ J/mol.K, and $T = 30^\circ$C. These parameters result in linear growth of the ASR gel with an approximately constant expansion rate as shown in Figure \ref{ASRfree-results}a. Moreover, the ASR gel Young's modulus is $E_{gel} = 10$ GPa, and its Poisson's ratio $\nu_{gel} = 0.2$, adapted from Leeman and Lura \cite{leemann2013modulus}. 

Contour of the damage variable $D$ at $450$ days, plotted in Figure \ref{ASRfree-geom-contour}b, indicates that the developed model is capable of capturing the random crack network distribution, which is characteristic of ASR free expansion. In order to investigate the effect of aggregates spatial distribution inside the simulated sample, three specimens with different internal aggregate realizations are modeled. All samples are generated with the same geometrical parameters, and the random positions of placed aggregates are only modified. Macroscopic expansion of the samples are measured along the three edges of the prisms and are plotted in Figure \ref{ASRfree-results}b. It can be observed that the internal aggregates configuration does not have considerable effect on the macroscopic expansion of the samples. Macroscopic expansion curves of the numerical simulations are averaged, and the average response is plotted along with the experimental measurement in Figure \ref{ASRfree-results}c, which shows a good match. The measured expansion curve in the experiment includes two distinct parts: the initial expansion with smaller rate, which is followed by a branch with higher expansion rate. Experiments reveal that the initial part corresponds to elastic expansion of the specimen and the primary stages of crack initiation in aggregates. The second branch is associated with the crack network development in aggregates and  propagation in paste \cite{dunant2009experimental}. It is interesting to see that the finite-element model is capable of reproducing the two branches of the expansion curve by employing a simple linear gel expansion. The red curve in Figure \ref{ASRfree-results}c shows the sample axial expansion when aggregates and paste are considered to behave elastically. Comparing this elastic response with the blue curve, in which damage behavior is taken into account, one can conclude that the major part of the macroscopic ASR expansion in concrete is due to crack network formation. For this specific simulation, $6\%$ of the final expansion at 450 days is due to elastic deformation, and  $94\%$ is associated with damage and crack opening. The total volume of the damaged finite elements in aggregates and paste are calculated during the simulation, and the evolution of the ratio of each quantity to the whole volume of the corresponding phase is plotted in Figure \ref{ASRfree-results}d. It is observed that the relative damaged volume of aggregates increases first, which is followed by the growth of the same quantity in paste as cracks that were initiated in aggregates propagate into paste. It should be considered that the values of the relative damaged volume curves of aggregates and paste do not represent the physical reality, and only the trend of the curves should be studied. In addition, it is observed that the aggregates damaged volume curve reaches a plateau after approximately 350 days, which means no further damage is generated in aggregates. On the other hand, although the rate of damage generation is reduced, additional damage is developed in the paste phase. The expansion curve obtained from the numerical simulation and the damaged volume curves presented in Figure \ref{ASRfree-results}d are normalized using their final value at 450 days and are plotted in Figure \ref{ASRfree-results}e. It is interesting to observe that the steepest points on the relative damaged volume curves, points $b$ and $c$ in Figure \ref{ASRfree-results}e, take place approximately at the time of the inflection point in the axial expansion curve, point $a$ in Figure \ref{ASRfree-results}e. This confirms that the initial branch of the expansion curve corresponds to elastic deformation of the sample, while the second branch is related to the damage propagation in aggregates and paste. 

\begin{figure}[t!]
  \centering 
  \includegraphics[width=0.8\textwidth]{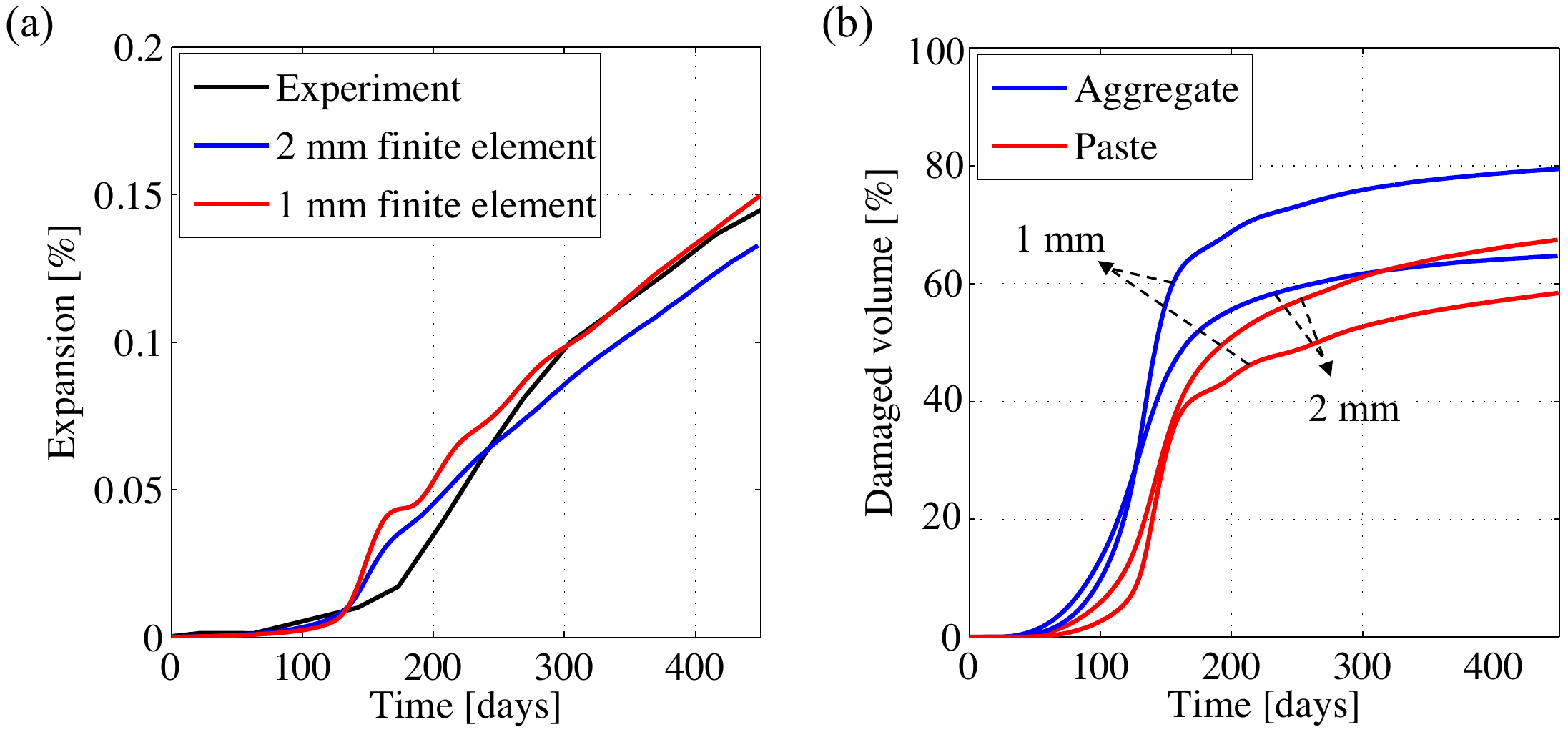}
  \caption{(a) Axial expansion of the prism discretized with 1 mm and 2 mm finite element size. (b) The evolution of damaged volume of aggregates and paste in the simulations. In both simulations, all parameters including $\overline{V}_{ga}$ are the same, which implies that the number of finite elements in aggregates converted to ASR gel is greater in the simulation with 1 mm finite element size compared to the one with 2 mm discretization size.}
  \label{FEsize-effect}
\end{figure}

To investigate the effect of finite element size, the prism is discretized using linear tetrahedral elements with 1 mm edge length. To ensure the dissipation of correct amount of fracture energy in aggregates and paste, the ultimate fracturing strain, after which damage variable $D=1$, is increased to 4.2\%. It should be considered that the total number of finite elements in aggregates that are converted to ASR gel is increased such that the volume fraction of gels $\overline{V}_{ga} = 2.5\%$ is maintained. ASR free expansion of the sample simulated with 1 mm finite element size is measured and plotted in Figure \ref{FEsize-effect}a, which shows a good agreement with the simulation results of the sample with 2 mm finite element size as well as the experimental data. Evolution of the relative damaged volume of aggregates and paste are plotted in Figure \ref{FEsize-effect}b. One can see that, while the macroscopic expansion of the samples with 1 and 2 mm discretization sizes are close at any given time, relative damaged volume of aggregates and paste are respectively higher and lower in the simulation with smaller element size. This is due to the fact that in order to attain a specific value of $\overline{V}_{ga}$ in the sample with finer mesh, higher number of finite elements should be converted to ASR gel, which results in a more diffused distribution of ASR product in aggregates. The trend of the curves related to the simulation with finer element discretization better resembles the experimental observations, as a higher volume of damage is observed in aggregates than paste, for instance the experiments of Ben Haha et al. \cite{haha2007relation}, in which damage evolution in concrete micro-structure is tracked via SEM imaging analysis. Another interesting study would be comparing 1 mm and 2 mm mesh sizes, while the number of gel pockets as well as the amount of gel volume percentage are kept constant. This implies that for the case of fine mesh, ASR gel size is equal to the one in the coarse mesh simulation, and each gel pocket is meshed with several finite elements. The presented results also indicate the importance of the simulated ASR gel size and emphasize on the necessity of accurate experimental measurements. It is worth repeating that the values of the damaged volumes in Figure \ref{FEsize-effect}b do not correspond to reality and should be studied to comprehend the trend of damage development and the influence of crucial parameters. In order to obtain a realistic estimate of the statistics of damage network developed during ASR, including the opening of cracks, their orientation, and the volume of crack clusters, a cohesive element approach should be taken into account. In this approach, cracks are developed on the interfaces between elastic elements, and their geometrical features can be directly linked to the experimental observations.

One should consider that the ASR experiments that are modeled in this paper are under free expansion condition. Experimental studies confirm the effect of confining stresses on concrete ASR expansion such as the work of Dunant and Scrivener. \cite{dunant2012effects}. To capture this effect in an accurate way, Mazars damage model employed in the current work should be extended to an orthotropic damage model, in which a damage tensor is defined in order to account for the material damage state in different directions. This means that when a crack is formed and opens along a certain direction, the corresponding finite element loses its stiffness in the direction perpendicular to the crack plan and maintains its resistance along in-plane ones. In ASR experiments under axial loads, the generated cracks are mostly aligned parallel to the loading axis. Thus, implementation of orthotropic damage model is necessary and is one of the future goals to pursue.

\subsection{Parameter sensitivity analysis}\label{sensytivity-analysis}
Developing such detailed meso-scale models requires additional information about the ASR gel mechanical properties, expansion rate of ASR gel, and the amount of reactive materials inside the aggregates. Determining these parameters for every concrete specimen in experiments is very intricate if not impossible. Therefore, the current meso-scale model relies on assumptions about these parameters. On the other hand, such mechanical models allow one to comprehend the effect of these influential factors on macroscopic ASR effect.

\begin{figure}[h!]
  \centering 
  \includegraphics[width=0.8\textwidth]{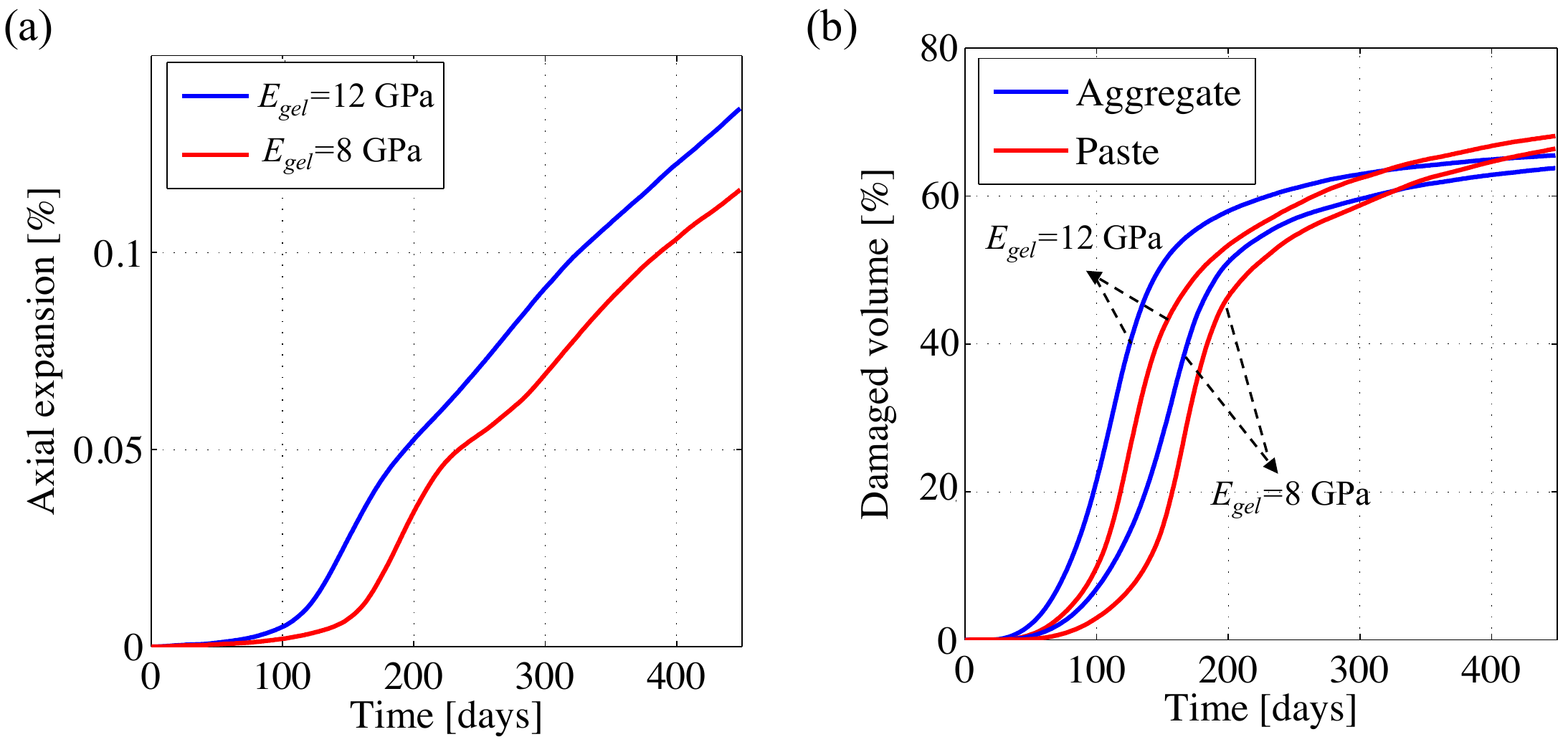}
  \caption{Effect of ASR gel Young's modulus on (a) axial expansion of the prism, and (b) evolution of damaged volume of aggregates and paste.}
  \label{ASRfree-gelElas}
\end{figure}

\subsubsection{ASR gel elastic modulus}
Research activities on evaluation of elastic and viscoelastic properties of ASR product have considerably increased during the last decade. Leeman and Lura \cite{leemann2013modulus} measured the elastic modulus of the ASR gel formed in the central part of four different aggregate types by nano-indentation and reported values in the range of $8$ to $12$ GPa. This range is associated with different chemical composition of ASR product in different kinds of aggregates. Effect of this property is studied by simulating the ASR free expansion experiment using the two margins of the reported range, while all other parameters are unaltered. Axial expansion versus time of the two simulations are plotted in Figure \ref{ASRfree-gelElas}a, and the evolution of damaged volume percentage of aggregates and paste are presented in Figure \ref{ASRfree-gelElas}b. One can observe that for higher value of $E_{gel}$, ASR expansion starts at earlier time and results in higher value of expansion at the end of the test. In addition, the damaged volume curves of aggregates and paste increase earlier for $E_{gel} = 12$ GPa, while they converge to approximately the same values for both $E_{gel}$ parameters. This implies that the final level of developed damaged volume in aggregates and paste are close in both simulations, and the damage network development is faster for the case of higher $E_{gel}$. This is due to the fact that the higher value of $E_{gel}$ gives rise to higher stresses in gel pockets and further higher tensile forces in aggregates. Thus, a stiffer gel pocket will require a lower value of strain to produce the same level of damage.

\begin{figure}[h!]
  \centering 
  \includegraphics[width=0.8\textwidth]{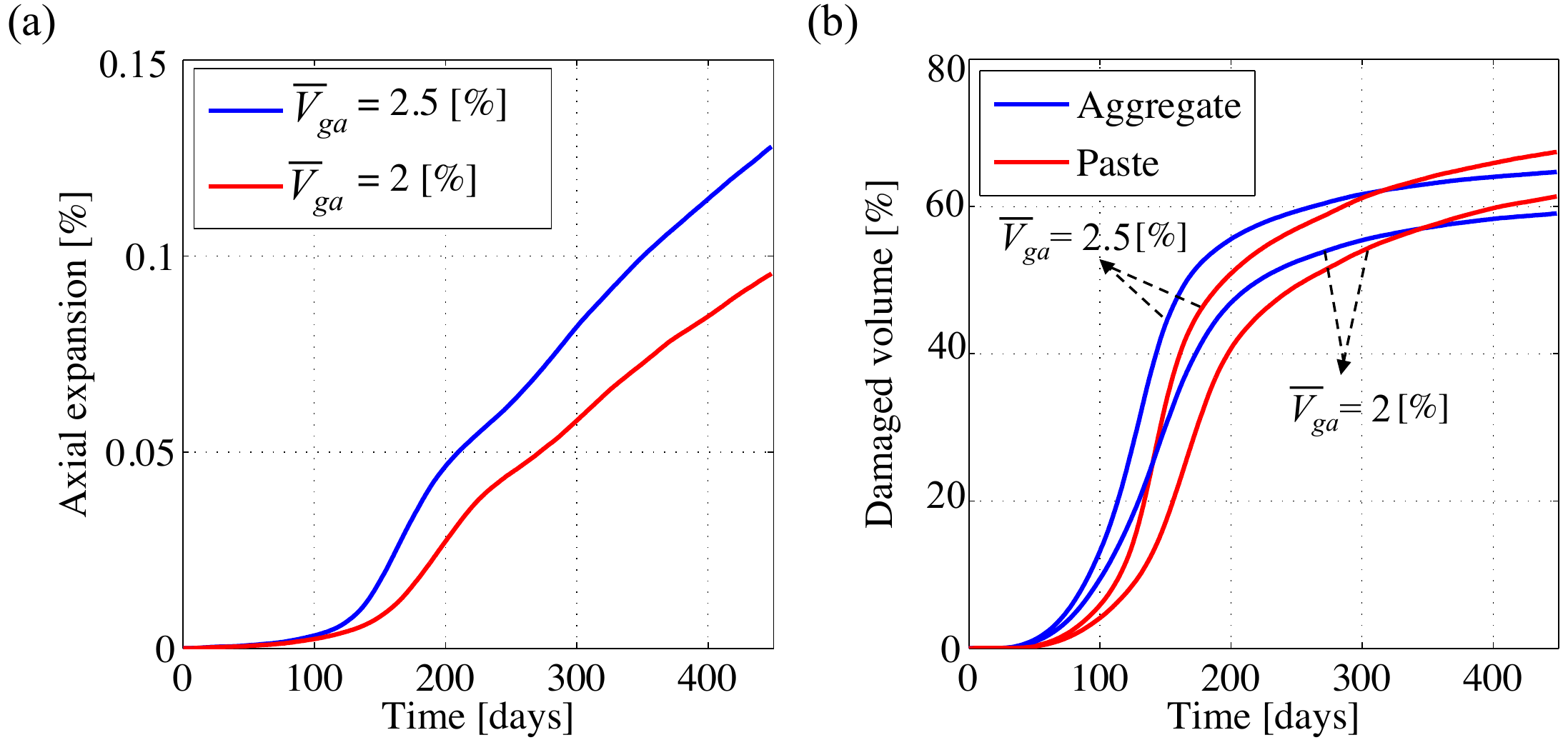}
  \caption{Effect of the amount of ASR gel on (a) axial expansion of the prism, and (b) evolution of damaged volume of aggregate and paste.}
  \label{ASRfree-gelperc}
\end{figure}

\subsubsection{ASR gel percentage}\label{ASRgel-perc}
The ratio of ASR gel volume to the volume of aggregates denoted by $\overline{V}_{ga}$ defines the number of finite elements in the model that are converted to ASR gel. Therefore, it represents the level of alkali content as well as aggregates reactivity in the experiments. Effect of the level of alkali content has been investigated in different experimental works \cite{shehata2000effect,sibbick1992threshold}, which show increase of ASR expansion with higher level of alkali content. This corresponds to the simulation of prism expansion with $\overline{V}_{ga} = 2$ and $2.5$\% depicted in Figure \ref{ASRfree-gelperc}a. Evidently, we observe in Figure \ref{ASRfree-gelperc}b, that the volume of the crack network in both aggregates and paste are higher for $\overline{V}_{ga} = 2.5$\%. The increase of the damaged volume generation rate, the steepest points on these curves, and their plateaus occur approximately at the same time for both $\overline{V}_{ga}$ values. This is consistent with the fact that the inflection points of both axial expansion curves presented in Figure \ref{ASRfree-gelperc}a occur simultaneously. 

\begin{figure}[h!]
  \centering 
  \includegraphics[width=0.8\textwidth]{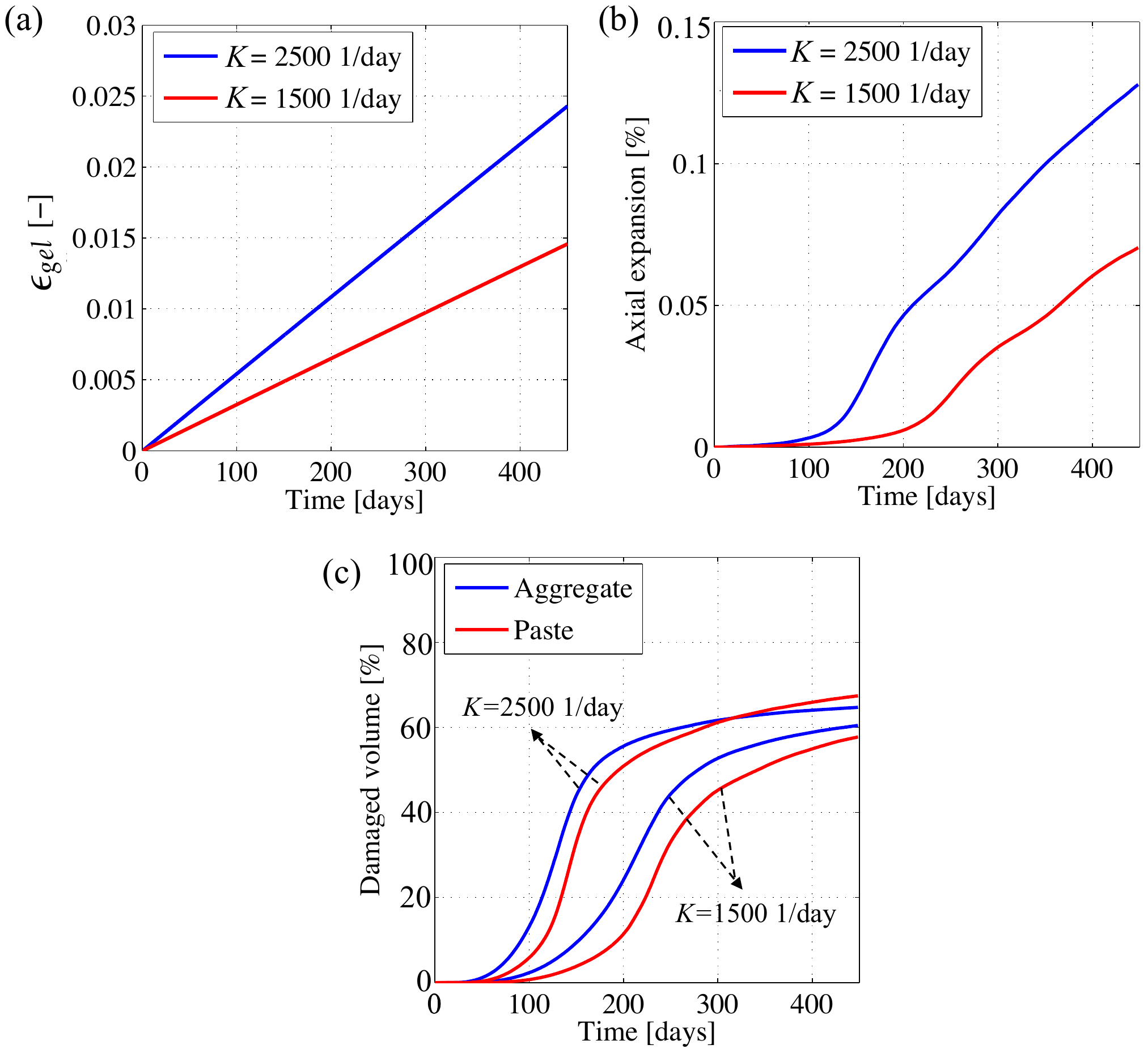}
  \caption{(a) ASR gel strain evolution with two different $K$ parameters. (b) Axial expansion of the prism due to ASR, and (c) evolution of damaged volume in aggregates and paste for the two different $K$ parameters.}
  \label{ASRfree-keffect}
\end{figure}

\begin{figure}[h!]
  \centering 
  \includegraphics[width=0.8\textwidth]{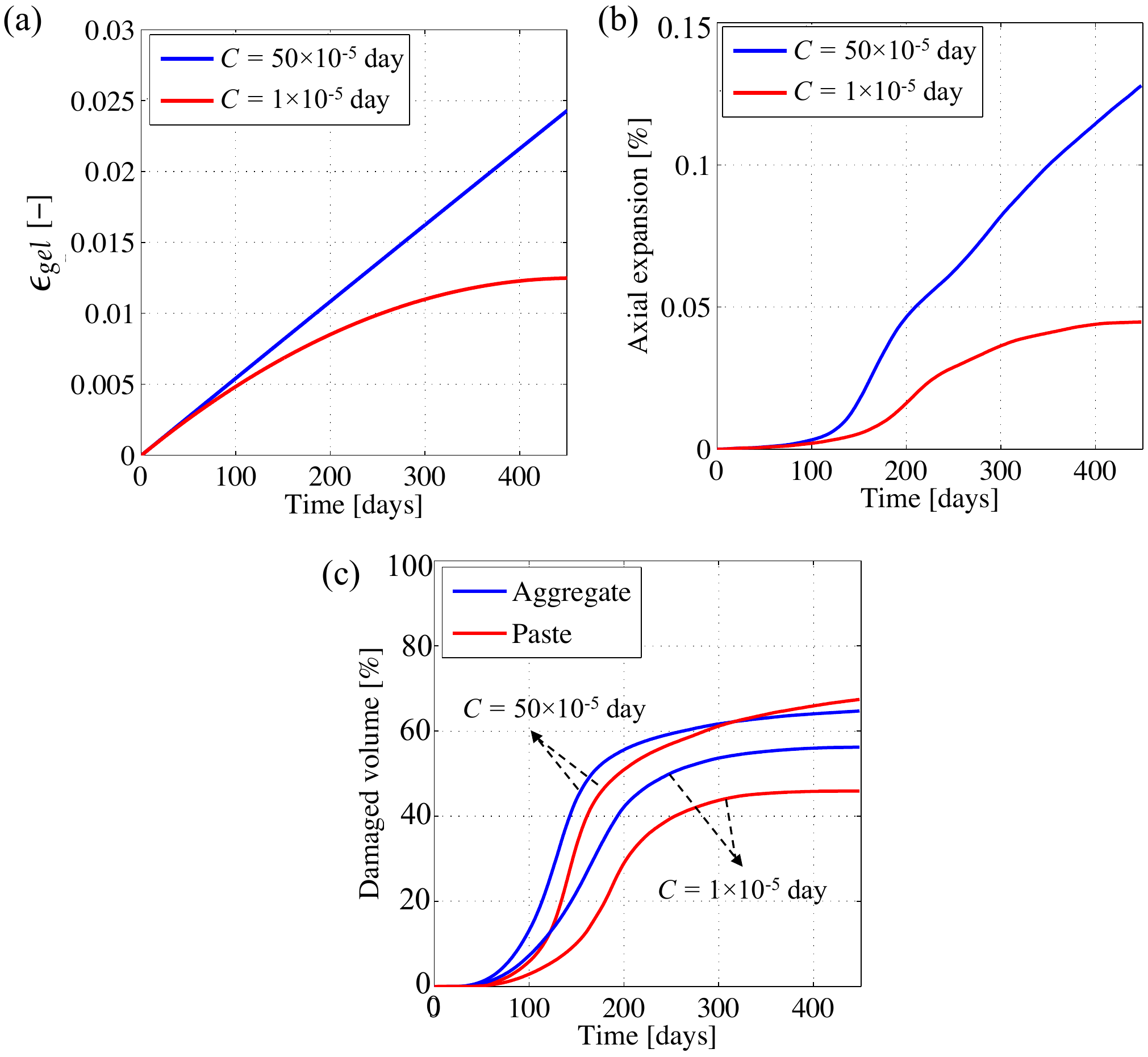}
  \caption{(a) ASR gel strain evolution with two different $C$ parameters. (b) Axial expansion of the prism due to ASR, and (c) evolution of damaged volume in aggregates and paste for the two different $C$ parameters.}
  \label{ASRfree-Ceffect}
\end{figure}

\subsubsection{ASR gel expansive behavior}
To study the effect of gel expansion rate, two values of $K = 1500$ and $2500$ 1/day are considered. Corresponding evolution of ASR gel strains $\epsilon_{gel}$ are depicted in Figure \ref{ASRfree-keffect}a, which shows linear gel expansion with higher expansion rate for $K = 2500$ 1/day. Increase of the ASR gel expansion rate shifts the inflection point of the expansion curve to earlier time and results in higher level of axial strain as shown in Figure \ref{ASRfree-keffect}b. This consequently yields to a faster growth of damaged volume in both aggregates and paste, while their final values seem to converge for both values of $K$, see Figure \ref{ASRfree-keffect}c. It is interesting to observe that with the shift of the inflection point on the curve of axial expansion, the steepest points of the damage evolution curves are adjusted as well. These results correspond to a similar study performed by Cuba Ramos et al. \cite{ramos2018hpc} in a two-dimensional setting. 

\begin{figure}[t!]
  \centering 
  \includegraphics[width=0.8\textwidth]{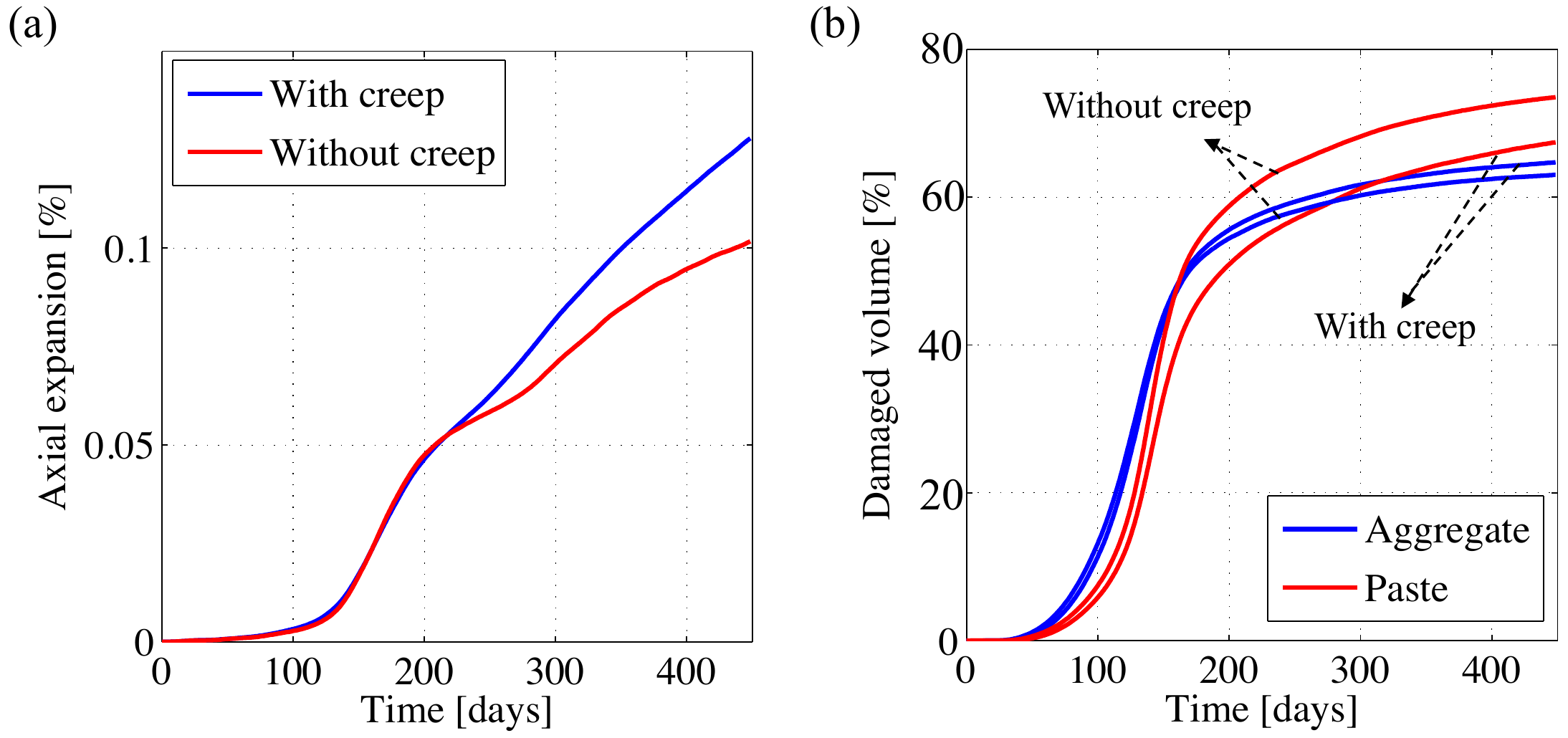}
  \caption{Effect of paste viscoelastic behavior on (a) axial expansion of the prism, and (b) evolution of damaged volume of aggregates and paste.}
  \label{ASRfree-creep-effect}
\end{figure}

Numerous experimental studies are available in the literature reporting measurements of ASR free expansion of concrete samples. Most of the previous studies have shown a plateau at the tail of the expansion curves, which can be due to different reasons. Among these reasons are the full consumption of alkali ions and alkali leaching from the sample. Although this is still an ongoing controversy, this aspect of ASR expansion can be captured using parameter $C$ in ASR gel expansive behavior formulated in Equation \ref{strain_rate-A}. ASR free expansion simulation is performed using $C = 50\times 10^{-5}$ and $C = 1\times 10^{-5}$ day, and the corresponding ASR gel expansion curves are presented in Figure \ref{ASRfree-Ceffect}a. The initial gel expansion rate of the two simulations are equal, while it gradually decays for the case of $C = 1\times 10^{-5}$ day. Considering Figure \ref{ASRfree-Ceffect}b, one can see that the macroscopic expansion curve becomes asymptotic with the reduction of the parameter $C$, while the inflection point of the expansion curves, which is tied to the initial gel expansion rate, occurs approximately at the same time. The final amount of damaged volume of aggregates and paste are evidently decreased with the reduction of gel expansion rate over time, as shown in Figure \ref{ASRfree-Ceffect}c. It can be seen that the final amount of damaged volume in paste is less than aggregates for the case of $C = 1\times 10^{-5}$. This is due to the fact that during ASR, the crack network initially generated in aggregates propagate into paste over the course of time. However, for the case of smaller $C$, this phase of the process is slowed down with the reduction of gel expansion rate. Therefore, a higher decrease of damaged volume takes place in paste in comparison with aggregates.

\subsubsection{Effect of viscoelasticity}
The effect of creep on ASR in concrete can be directly examined in the current model. The simulation of the prism ASR expansion is repeated by excluding viscoelastic properties from the mechanical behavior of paste. It is shown that creep has negligible influence on the prism free expansion in the early stages, and minor effect can be observed in later time of the experiment, when creep strain is pronounced in finite elements subjected to tensile stresses, see Figure \ref{ASRfree-creep-effect}a. Since this effect is dependent on the distribution of internal stresses in the paste phase, it is expected that with the increase of modeling resolution and convergence of the size of ASR product to reality, the difference between the two curves would diminish. On the other hand, paste viscoelasticity substantially decreases the volume of the developed damage network in paste, while it does not affect this quantity in aggregates, as shown in Figure \ref{ASRfree-creep-effect}b. These results are reasonable as paste viscoelasticity relaxes the stresses generated in paste due to aggregates expansion and reduces the level of damage in this phase. 

\begin{figure}[t!]
  \centering 
  \includegraphics[width=0.8\textwidth]{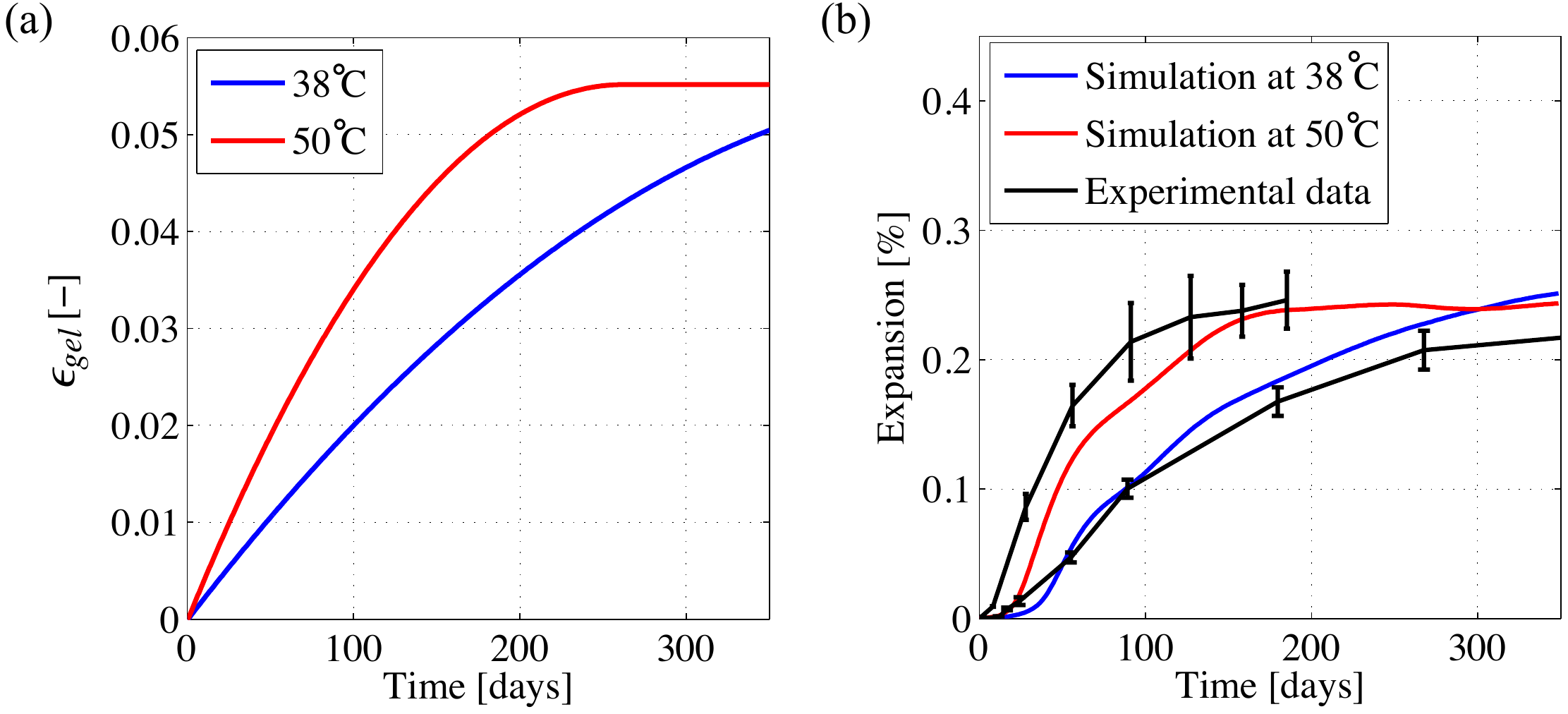}
  \caption{(a) ASR gel expansion at two different temperatures. (b) Average of prismatic sample expansion along three edges versus time in comparison with experiments reported by Gautam et al. \cite{gautam2017effect}.}
  \label{ASRfree-temp-effect}
\end{figure}

\subsubsection{Effect of temperature}
In order to study the effect of temperature on ASR process, the experiments presented by Gautam et al. \cite{gautam2017effect} are considered in numerical simulations. In these experiments, duration of ASR test is reduced by increasing the reaction rate through elevating ambient temperature. Concrete Prism Test (CPT) is performed on prismatic samples with $75\times75\times285$ mm under constant temperature of 38$^\circ$C and 50$^\circ$C. Relative humidity of the samples are maintained above 95\% during the experiments. In numerical simulations, prismatic samples of $75\times75\times140$ mm are modeled to reduce the computational cost. Minimum aggregate size in the simulation is 4 mm, and the maximum aggregate size is 19 mm. Whole specimen is discretized with 2 mm linear tetrahedral finite elements, and the mechanical properties of all constituents are identical to the ones presented in Section \ref{ASR-free}. ASR gel expansion parameters, $C$ and $K$, and the percentage of ASR product $\overline{V}_{ga}$ are calibrated for the experimental measurements at 38$^\circ$C, and the calibrated parameters are: $C = 17\times 10^{-6}$ day; $K = 6500$ 1/day; and $\overline{V}_{ga} = 2.8\%$. Using these parameters, ASR gel strain evolution at 38$^\circ$C is shown in blue in Figure \ref{ASRfree-temp-effect}a. The corresponding specimen expansion measurement during 350 days is plotted in Figure \ref{ASRfree-temp-effect}b along with the related experimental data, which shows a good match between the laboratory measurements and simulation results. Using the calibrated parameters, temperature parameter $T$ in Equations \ref{strain_rate} and \ref{strain_rate-A} is increased to 50$^\circ$C, while all other parameters are kept unaltered. Temporal variation of gel strain at 50$^\circ$C is plotted in Figure \ref{ASRfree-temp-effect}a in red. One can see that initial gel expansion rate is increased, which gradually decays and reaches a plateau. Numerical simulation results of the specimen expansion at the elevated temperature are depicted in Figure \ref{ASRfree-temp-effect}b in red in comparison with the corresponding experimental data recorded for 187 days. One can see that the finite element model predicts well the laboratory measurements at the increased temperature, also referred to as Accelerated Concrete Prism Test (ACPT), both in terms of the initial growth of the expansion rate as well as its asymptotic tail. The error bars in the experimental curves represent the scatter of the data, and the simulation results are the average of numerical simulation of three samples with different internal aggregate configurations. Accounting for a proper accelerating ratio due to 12$^\circ$C temperature increase, Gautam et al. \cite{gautam2017effect} has shown that ACPT can reproduce the CPT results at a higher speed. They have shown that through an appropriate time scaling, the expansion trend of both tests are identical and the final value of specimen expansions are approximately equal at both ambient temperatures. The later aspect is interestingly verified by the numerical simulations.

\section{Conclusions}	
In this paper, a three-dimensional meso-scale finite element model is proposed to study the effect of ASR in concrete. Coarse aggregate pieces, cement paste, and ASR product are explicitly modeled, which allows to investigate the effect of different influential parameters. It is shown that the ASR free expansion of a prismatic concrete sample in laboratory experiments can be reasonably reproduced by the numerical simulations. The initial elastic expansion and the increase of the expansion rate with the associated inflection point can be captured with linear growth of ASR gel. Such characteristics in ASR expansion curves are achieved in macroscopic models by presuming an ``S" shape for gel expansion curve, which considers a slow expansion kinetic in the initial growth phase. In addition, it is shown that major portion of specimen expansion is due to the development of crack network, and a small fraction is related to the elastic expansion. Furthermore, the developed model is employed to examine the effect of ASR gel elastic modulus, its rate of expansion, and the amount of reactive materials. It is shown that the increase of the first two factors, although resulting in higher level of expansion, approximately develops the same damaged volume in both aggregates and paste. This indicates that a higher level of crack opening is obtained in the specimen when $E_{gel}$ and $\dot{\epsilon}_{gel}$ are increased. Moreover, stiffer and more expansive ASR gel results in a higher level of damage already at early stages of the reaction. On the other hand, increasing the amount of reactive material $\overline{V}_{ag}$ gives rise to a higher amount of damaged volume in both phases, while it does not affect the ASR kinetics. This observation can be justified by associating the number of formed cracks solely to the number of gel pockets placed into the model, which emphasizes the importance of explicitly modeling the gel pockets in the mechanical model. Moreover, creep effect on reduction of the amount of damage in paste is well captured through comparing prism simulations with and without viscoelastic characteristics in paste. Finally, temperature effect on accelerating the ASR expansion test is reproduced in accordance with the experimental measurements.

Regarding the future extension of the model, there are several interesting research avenues that we can pursue based on the developed model. Among these research directions are the influence of confining stresses, aggregates size and shape effects, effect of varying ambient conditions by considering diffusion of temperature and humidity inside the specimen, and the effect of viscoelastic nature of the ASR product.
\vspace{5mm}\\
\noindent\textbf{ACKNOWLEDGMENTS} \\
This research is supported by the SNF Sinergia project ``Alkali-silica reaction in concrete (ASR)'', grant number CRSII5\_17108.

\bibliographystyle{unsrt}
\bibliography{citations}

\begin{appendices}
\numberwithin{equation}{section}

\section{Viscoelastic behavior calculations in explicit time integration}\label{visc-explicit}
For a given time step $n$ from $t_n$ to $t_{n+1} = t_n + \Delta t$, the strain increment $\Delta \boldsymbol{\epsilon}_n$ is known. Using time mapping, one can calculate real physical time step $\Delta t^r$ through $\Delta t^r = \Delta t \times T^r/T^{sim}$. $T^r$ is the real physical final time of the experiment, and $T^{sim}$ is the final time of the numerical simulation. Current physical time $t_{n+1}^r$ is also calculated through $t_{n+1}^r = (n+1) \times \Delta t^r$. The real physical time step and current time are then used in the calculation of viscoelastic constitutive equations. Two factors $\beta_{\mu} = \text{e}^{-\Delta t^r/\tau_\mu}$ and $\lambda_{\mu} = (1-\beta_{\mu })\frac{\tau_\mu}{\Delta t^r}$ are calculated for each viscous unit, which do not change through the analysis by considering a constant time step. Incremental modulus ${E}_n$ for time step $n$ is then calculated as follows
\begin{equation}\label{incr-modulus}
{E}_n = \bigg(\frac{1}{E_0} + \frac{1}{v_{n+\frac{1}{2}}} \sum_{\mu=1}^N \frac{1-\lambda_{\mu}}{E_\mu^\infty} \bigg)^{-1}
\end{equation}
in which $v_{n+\frac{1}{2}} = v(t_n^r+\frac{1}{2}\Delta_t^r)$. The creep strain increment can be calculated through
\begin{equation}\label{creep-str-incr}
\Delta \boldsymbol{\epsilon}_n^{cr} = \frac{\bold{C}_\nu}{v_{n+\frac{1}{2}}} \sum_{\mu=1}^N \frac{1-\beta_\mu}{E_{\mu}^{\infty}}\boldsymbol{\sigma}_{\mu,n}
\end{equation}
in which 
\begin{equation}\label{cv}
\bold{C}_\nu=
  \begin{bmatrix}
    1 & -\nu & -\nu & 0 & 0 & 0 \\
    -\nu & 1 & -\nu & 0 & 0 & 0 \\
    -\nu & -\nu & 1 & 0 & 0 & 0 \\
    0 & 0 & 0 & 2(1+\nu) & 0 & 0\\
    0 & 0 & 0 & 0 & 2(1+\nu) & 0\\
    0 & 0 & 0 & 0 & 0 & 2(1+\nu)\\
  \end{bmatrix}
\end{equation}

$\boldsymbol{\sigma}_{\mu,n}$ is the viscous stress in the rheological unit $\mu$ at time $t_n$, which is an internal variable and is updated and saved for the calculations in the next time step as
\begin{equation}\label{sigma_v_n+1}
\boldsymbol{\sigma}_{\mu,n+1} = \lambda_\mu \Delta \boldsymbol{\sigma_n} + \beta_\mu \boldsymbol{\sigma}_{\mu,n}
\end{equation}

$\Delta \boldsymbol{\sigma_n}$ in Equation \ref{sigma_v_n+1} reads
\begin{equation}\label{stress-incr}
\Delta \boldsymbol{\sigma_n} = {E}_n \bold{D}_\nu (\Delta \boldsymbol{\epsilon}_n - \Delta \boldsymbol{\epsilon}_n^{cr})
\end{equation}
where $\bold{D}_\nu = \bold{C}_\nu^{-1}$, and $\Delta \boldsymbol{\epsilon}_n^{els} = (\Delta \boldsymbol{\epsilon}_n - \Delta \boldsymbol{\epsilon}_n^{cr})$ is the elastic strain increment. Furthermore, the elastic stress tensor is calculated by
\begin{equation}\label{stress-total}
\boldsymbol{\sigma}_{n+1}^{els} = {E}_n \bold{D}_\nu (\boldsymbol{\epsilon}_{n+1} - \boldsymbol{\epsilon}_{n+1}^{cr})
\end{equation}
in which $\boldsymbol{\epsilon}_{n+1}^{els} = (\boldsymbol{\epsilon}_{n+1} - \boldsymbol{\epsilon}_{n+1}^{cr})$ is the elastic strain tensor, which is used in the calculation of damage parameter $D$ as explained in Appendix \ref{mazars-explicit}.

\section{Mazars damage law calculations in explicit time integration}\label{mazars-explicit}
At any given time during the simulation $t_{n+1}$, elastic strain tensor calculated as $\boldsymbol{\epsilon}_{n+1}^{els} = (\boldsymbol{\epsilon}_{n+1} - \boldsymbol{\epsilon}_{n+1}^{cr})$ and the corresponding elastic stress tensor $\boldsymbol{\sigma}_{n+1}^{els}$ calculated in Equation \ref{stress-total} are the inputs to the Mazars damage model. One should consider that for the case of aggregates, $\boldsymbol{\epsilon}_{n+1}^{cr}=0$, and the elastic strain tensor is equal to the strain tensor. $\boldsymbol{\epsilon}_{n+1}^{els}$ is transformed to its principal coordinate system leading to three principal strains $\epsilon_{I}$ where $I=1,2,3$. One should consider that the subscript $n+1$ and the elasticity superscript are dropped from the principal stress and strain quantities in the rest of this section for the sake of simplicity. Next, the positive principal strains $\left\langle \epsilon_{I} \right\rangle _{+}$ are selected to calculate the effective elastic strain $\tilde{\epsilon}$ using Equation \ref{effective-strain}. If ($\tilde{\epsilon} > k_0$), then damage variables in tension $D_t$ and compression $D_c$ are calculated using Equation \ref{damage-decompos}. To compute the total damage variable $D$, $\alpha_t$ and $\alpha_c$ should be calculated using Equation \ref{alpha_tc}. To compute $\epsilon_{I}^t$ in $\alpha_t$ formula in Equation \ref{alpha_tc}, principal stress tensor $\boldsymbol{\sigma}_{p}$ is calculated using the principal strain tensor  $\boldsymbol{\epsilon}_{p}$ through
\begin{equation}\label{princ-stress}
\boldsymbol{\sigma}_{p} = E_n \bold{D}_\nu \boldsymbol{\epsilon}_{p} 
\end{equation}
where in Voigt notation, the first three components of $\boldsymbol{\epsilon}_{p}$ are the principal strains $\epsilon_{I}$ with $I=1,2,3$, and the other three components are zero. Having $\boldsymbol{\sigma}_{p}$ calculated, the positive principal stresses $\left\langle \sigma_{I} \right\rangle _{+}$ are then used to form the tensile principal stress tensor $\boldsymbol{\sigma}_{p}^t$, which is employed for the calculation of positive principal strain tensor $\boldsymbol{\epsilon}_{p}^t$ as follows
\begin{equation}\label{princ-strain-pos}
\boldsymbol{\epsilon}_{p}^t = E_n^{-1} \bold{C}_\nu \boldsymbol{\sigma}_{p}^t
\end{equation}
in which $\epsilon_{I}^t$ with $I=1,2,3$ are the first three components of $\boldsymbol{\epsilon}_{p}^t$. Using the calculated $\epsilon_{I}^t$ in Equation \ref{alpha_tc}, $\alpha_t$ and $\alpha_c = 1 - \alpha_t$ are computed. Finally, damage variable $D$ is calculated using Equation \ref{damage-decompos}, and the stress tensor at the end of the current time step is computed using elastic stress tensor calculated in Equation \ref{stress-total} as
\begin{equation}\label{stress-total-damage}
\boldsymbol{\sigma}_{n+1} = (1-D) {E}_n \bold{D}_\nu (\boldsymbol{\epsilon}_{n+1} - \boldsymbol{\epsilon}_{n+1}^{cr})
\end{equation}

\end{appendices}

\end{document}